\documentclass[11pt,a4paper]{article}
\pdfoutput=1
\usepackage{graphicx}
\usepackage{amssymb,amsmath,mathtools,mathrsfs} 
\usepackage[normalem]{ulem}
\usepackage{enumerate,todonotes} 
\usepackage{jcappub}
\usepackage{placeins}
\usepackage{multirow}
\usepackage{hyperref}
\usepackage{float,todonotes}
\usepackage{setspace}
\usepackage{graphics}
\usepackage{fullpage}
\usepackage{epstopdf}
\usepackage{aas_macros}
\usepackage{xfrac}
\usepackage{tabularx,subfig}
\usepackage[export]{adjustbox}
\usepackage[utf8]{inputenc}
\usepackage{arydshln}

\newcommand{\be}{\begin{equation}}
\newcommand{\ee}{\end{equation}}
\newcommand{\dof}{{\it dof}}
\newcommand{\dofs}{{\it dof}}

\newcommand{\nin}{\noindent}
\newcommand{\comment}[1]{}

\newcommand\Tstrut{\rule{0pt}{3ex}}

\makeindex
\begin{document}
\hypersetup{pageanchor=false} 
\title{Cosmological parameter constraints\\ for Horndeski scalar-tensor gravity}

\author[a,b]{Johannes Noller,}
\author[b,c]{Andrina Nicola}
\affiliation[a]{Institute for Theoretical Studies, ETH Z\"urich, Clausiusstrasse 47, 8092 Z\"urich, Switzerland}
\affiliation[b]{Institute for Particle Physics and Astrophysics, ETH Z\"urich, 8093 Z\"urich, Switzerland}
\affiliation[c]{Department of Astrophysical Sciences, Princeton University, Princeton, NJ 08544, USA}

\emailAdd{johannes.noller@eth-its.ethz.ch}
\emailAdd{anicola@astro.princeton.edu}

\abstract{
We present new cosmological parameter constraints for general Horndeski scalar-tensor theories, using CMB, redshift space distortion, matter power spectrum and BAO measurements from the Planck, SDSS/BOSS and 6dF surveys. 
We focus on theories with cosmological gravitational waves propagating at the speed of light, $c_{\rm GW} = c$,
implementing and discussing several previously unaccounted for aspects in the constraint derivation for such theories, that qualitatively affect the resulting constraints. 
In order to ensure our conclusions are robust, we compare results for three different parametrisations of the free functions in Horndeski scalar-tensor theories, identifying several parametrisation-independent features of the constraints. 
We also consider models, where $c_{\rm GW} \neq c$ in cosmological settings (still allowed after GW170817 for frequency-dependent $c_{\rm GW}$) and show how this affects cosmological parameter constraints.
}


\maketitle


\section{Introduction}\label{sec:intro}

General Relativity (GR) at present remains firmly entrenched as a cornerstone of the cosmological standard model. 
Nevertheless we do know that GR is not the final answer. It is an effective theory that breaks down at Planck energies, is not geodesically complete and is plagued by fundamental problems, most notably the (old) cosmological constant problem. Since GR is the unique consistent theory of a massless spin-2 field (assuming Lorentz invariance), any attempt to modify or extend it in order to address one of these shortcomings will generically introduce new gravitational (light) degrees of freedom (dof). As such, one ought to be on the lookout for any signs of such new \dof{}, not just because their detection would revolutionise our understanding of gravity, but also since (in the absence of a detection) this is the most stringent way to test and put constraints on GR itself. With the increasing precision of current and upcoming data, cosmology provides an ideal testbed for the presence of such new gravitational \dofs{}.   

Before contrasting theory with data, one ought to make a choice on how to parametrise potential deviations from GR. Horndeski scalar-tensor theories \cite{Horndeski:1974wa,Deffayet:2011gz} have been the primary workhorse of modified gravity in recent times. They encompass and provide a minimal extension of GR in the sense that only one new single \dof{} is introduced, yet this is done with a set of theoretical constraints (notably Lorentz invariance and the absence of higher-derivative ghosts) that ensure one is working with a fundamentally sound theory space. 
As Horndeski scalar-tensor theories include the vast majority of scalar-tensor theories considered in the literature, but their theory space is nevertheless described by only a few interaction terms in the Lagrangian, these theories provide a simultaneously rich and well-constrained setup in which to place constraints on deviations from GR and the emergence of new gravitational \dofs{}.

In this paper, we therefore take Horndeski scalar-tensor theories and constrain them using data from several cosmological probes, specifically the cosmic microwave background (CMB) \cite{Planck-Collaboration:2016ae, Planck-Collaboration:2016aa}, baryon acoustic oscillations (BAOs) \cite{Anderson:2014, Ross:2015}, redshift space distortions (RSDs) \cite{Beutler:2012, Samushia:2014} and the matter power spectrum \cite{Tegmark:2006}.  In the process, we especially focus on the following four questions: 
I) What are the cosmological parameter constraints for theories, where gravitational waves propagate at the speed of light, $c_{\rm GW} = c$?
II) What are the corresponding constraints for theories, where the speed of gravitational waves is allowed to differ from that of light? What are the cosmological constraints on $c_{\rm GW}$ then and how does this additional freedom impact constraints on other parameters? 
III) Horndeski theories are spanned by four free functions, each in principle requiring an infinite number of parameters to be fully specified. One therefore needs to choose a more restrictive and specific ansatz for these functions in order to efficiently extract cosmological constraints. What parametrisation(s) should one choose and what cosmological constraints are robust under a change of parametrisation? 
IV) What datasets provide the most stringent constraints? Do they preferentially point towards specific modified gravity theories and are there (hints of) deviations from GR? What additional theoretical priors should one impose?
\\

\nin{\it Outline}: This paper is organised as follows. In section \ref{sec:horn}, we recap Horndeski scalar-tensor theories, their linearly perturbed action and how these can be significantly simplified by requiring gravitational waves to propagate at the speed of light, $c_{\rm GW} = c$ (as extensively discussed in the wake of GW170817 and GRB 170817A). In section \ref{sec:param}, we then consider different parametrisations for the remaining functional freedom and discuss them alongside additional theoretical (stability) constraints. This is followed by an overview of the different data sets used to extract cosmological parameter constraints in section \ref{sec:data}. In section \ref{sec:constraints} we then present the constraints for theories with $c_{\rm GW} = c$, discuss what essential aspects drive the constraints and how to best interpret the results, what constraints are robust under changes of parametrisations and what they mean for dark energy/modified gravity theories. 
In section \ref{sec:ct}, we recap cosmologically relevant caveats in the argument that infers $c_{\rm GW} = c$ from GW170817 and GRB 170817A, which imply that $c_{\rm GW} \neq c$ is still a valid setup on cosmological scales. We discuss how constraints change, if the speed of gravitational waves is allowed to vary, and present the corresponding Monte Carlo Markov Chain (MCMC) analysis.  
Finally, we conclude in section \ref{sec:conc} and provide further details in the appendices.  
\\

\nin{\it Notation and conventions}: 
Since we will be considering scalar-tensor theories, the principal ingredients will be a tensor $g_{\mu\nu}$ and a scalar $\phi$. The covariant derivative associated with $g_{\mu\nu}$ is $\nabla_\mu$ and we will introduce the shorthand $\Phi_{\mu\nu} \equiv \nabla_\mu\nabla_\nu \phi$. Finally, angular brackets denote taking the trace, so e.g. $[\Phi] = \Phi_\mu{}^\mu$ and $[\Phi^2] = \Phi_{\mu\nu}\Phi^{\mu\nu}$.

\section{Horndeski gravity} \label{sec:horn}

Here we briefly summarise the essential features of Horndeski scalar-tensor theories in a gravitational context, how they are defined, what free functions span the associated theory space and how these can be efficiently captured at the level of the linearised action. 

\subsection{Horndeski scalar tensor-theories}

The most general Lorentz-invariant scalar-tensor action that gives rise to second-order equations of motion (and is consequently free of an Ostrogradski-ghost instability by default), is Horndeski gravity \cite{Horndeski:1974wa,Deffayet:2009wt}:
\begin{eqnarray}\label{Horndeski_action}
S=\int d^4x \sqrt{-g}\left\{\sum_{i=2}^5{\cal L}_i[\phi,g_{\mu\nu}]\right\},
\end{eqnarray}
where the ${\cal L}_i$ are scalar-tensor Lagrangians given by:
\begin{eqnarray}
{\cal L}_2&=& G_2(\phi,X) , \nonumber \\
{\cal L}_3&=& -G_3(\phi,X)[\Phi] , \nonumber \\
{\cal L}_4&=&  G_4(\phi,X)R+G_{4,X}(\phi,X)\left([\Phi]^2-[\Phi^2]\right) , \nonumber \\
{\cal L}_5&=& G_5(\phi,X)G_{\mu\nu}\Phi^{\mu\nu}
-\frac{1}{6}G_{5,X}(\phi,X)\left( [\Phi]^3
-3[\Phi^2][\Phi] +2[\Phi^3]
\right)  \,. \label{Horndeski_lagrangians}
\end{eqnarray}
Four free functions ($G_{2}, G_3, G_4, G_5$) therefore completely characterise this theory. The $G_i$ are functions of a scalar field $\phi$ and its derivative via $X \equiv -\tfrac{1}{2}\nabla_\mu \phi \nabla^\mu \phi$.\footnote{The fact that the Lagrangian only depends on the first derivative via $X$ is a consequence of Lorentz invariance.} Finally, $G_{i,\phi}$ and $G_{i,X}$ denote the partial derivatives of the $G_i$ with respect to $\phi$ and $X$ respectively.

In the aftermath of the near simultaneous detections of GW170817 and GRB 170817A \cite{PhysRevLett.119.161101,2041-8205-848-2-L14,2041-8205-848-2-L15,2041-8205-848-2-L13,2041-8205-848-2-L12} it was shown in \cite{Baker:2017hug,Creminelli:2017sry,Sakstein:2017xjx,Ezquiaga:2017ekz} that imposing $c_{\rm GW} = c$ in a cosmological context significantly reduces the full Horndeski theory space \eqref{Horndeski_lagrangians}, namely by eliminating $G_5$ and $G_{4,X}$, as we will discuss in the next subsection.
Note that we will re-visit this argument in section \ref{sec:ct}, where we recap why extrapolating the measurement of $c_{\rm GW} = c$ from GW170817 and GRB 170817A to a cosmological context requires additional non-trivial assumptions and we discuss varying $c_{\rm GW}$ models in setups where these assumptions do not hold.
Putting this issue aside for the time being, imposing $c_{\rm GW} = c$ in a cosmological context reduces Horndeski theory to
\begin{align} \label{Horndeski_simple}
S=\int d^4x \sqrt{-g}\left\{ G_2(\phi,X) -G_3(\phi,X)[\Phi] + G_4(\phi) R\right\},
\end{align}
where there are now only three free functions left ($G_{2}, G_3, G_4$) and $G_4$ is a function of $\phi$ only.\footnote{Note that imposing $c_{\rm GW} = c$ for cosmology only enforces $G_{5,X} = 0$, if the scalar \dof{} affects the cosmological background evolution, as it certainly should if it is at all related to dark energy/modified gravity. However, this does mean, that for theories where the scalar is sufficiently suppressed and does not affect cosmological evolution, $G_{5,X} = 0$ may be consistently violated \cite{Sakstein:2017xjx}, as is the case for Einstein-dilaton-Gauss-Bonnet theories (EdGB), that are of interest e.g. in strong gravity phenomenology.}
For previous related work on $c_{\rm GW} = c$ constraints see \cite{Amendola:2012ky,Amendola:2014wma,Linder:2014fna,Raveri:2014eea,Saltas:2014dha,Lombriser:2015sxa,Lombriser:2016yzn,Jimenez:2015bwa,Bettoni:2016mij,Sawicki:2016klv}.

\subsection{Linearised perturbations}

With a cosmological setting in mind, the general Horndeski action \eqref{Horndeski_action} can be expanded around a spatially flat homogeneous and isotropic background. Doing so to quadratic order in the (linear) perturbations yields the linearised dynamics of \cite{Bellini:2014fua,Gleyzes:2013ooa,Lagos:2016wyv,Lagos:2017hdr} -- also see \cite{Gubitosi:2012hu,Bloomfield:2012ff}. Here we will not repeat the derivation of the associated action, but instead note that the dynamics of linearised perturbations is completely controlled by four functions \cite{Bellini:2014fua}: They are the effective Planck mass $M_S$ and its running $\alpha_M$, the kineticity $\alpha_K$ that contributes to the kinetic energy of scalar perturbations, the braiding $\alpha_B$ that quantifies the strength of kinetic mixing between scalar and tensor perturbations, and the tensor speed excess $\alpha_T$, which is related to the speed of sound of tensor perturbations $c_T$ via $c_{\rm GW}^2 = 1 + \alpha_T$. In terms of the model functions $G_i$ these are given by \cite{Bellini:2014fua}
\begin{eqnarray}
M^2_S&\equiv&2\left(G_4-2XG_{4,X}+XG_{5,\phi}-{\dot \phi}HXG_{5,X}\right) , \nonumber \\
HM^2_S\hat{\alpha}_M&\equiv&\frac{d}{dt}M^2_S , \nonumber \\
HM^2_S\hat{\alpha}_B&\equiv&2\dot{\phi}\left(XG_{3,X}-G_{4,\phi}-2XG_{4,\phi X}\right) \nonumber \\ & &
+8XH\left(G_{4,X}+2XG_{4,XX}-G_{5,\phi}-XG_{5,\phi X}\right) \nonumber \\ & &
+2\dot{\phi}XH^2\left(3G_{5,X}+2XG_{5,XX}\right) \nonumber , \\ 
M^2_S\hat{\alpha}_T&\equiv&2X\left[2G_{4,X}-2G_{5,\phi}-\left(\ddot{\phi}-\dot{\phi}H\right)G_{5,X}\right] \,,
\label{alphadef}
\end{eqnarray}
where all the $G_i$ as well as $\phi$ and $X$ are evaluated for the background configuration. We further use the shorthand $G_i \equiv G_i(\phi,X)$ and refer to \cite{Bellini:2014fua} for the (lengthy) expression for $\alpha_K$.

These expressions greatly simplify when we specialise to the restricted Horndeski theories \eqref{Horndeski_simple} with luminally propagating gravitational waves. In that case, one trivially obtains $\alpha_T = 0$ and, collecting results for the $\alpha_i$, we obtain
\begin{eqnarray}
M^2_S&=&2 G_4, \nonumber \\
HM^2_S\hat{\alpha}_M&=&\frac{d}{dt}M^2_S , \nonumber \\
H^2M^2_S\hat{\alpha}_K&=&2X\left(G_{2,X}+2XG_{2,XX}-2G_{3,\phi}-2XG_{3,\phi X}\right)
+12\dot{\phi}XH\left(G_{3,X}+XG_{3,XX}\right) \nonumber, \\
HM^2_S\hat{\alpha}_B&=&2\dot{\phi}\left(XG_{3,X}-G_{4,\phi}\right), \nonumber \\ 
\hat{\alpha}_T&=& 0 \,.
\end{eqnarray}
Note that, as before, all parameters are determined in terms of the three free functions ($G_{2}, G_3, G_4$), where $G_4$ is a function of $\phi$ only and ($G_{2}, G_3$) can be functions of both $\phi$ and $X$.

\section{Parameterisations and stability conditions} \label{sec:param}

The $\alpha_i$ functions discussed above map the functional freedom from the full Horndeski action (captured by $G_{2}, G_3, G_4, G_5$) into their physically relevant combinations at the level of the linearised action. 
In order to extract meaningful constraints for these functions, it is necessary to reduce their inherent functional freedom by using some parametrised form for these functions.
Indeed this is also the approach implemented in state-of-the-art Einstein-Boltzmann solvers for Horndeski theories, such as hi\_class \cite{hiclass} and EFTCAMB \cite{EFTCAMB1}. The purpose of such parametrisations is to capture the dark energy evolution to reasonable accuracy in the late-universe. While naturally most simple parametrisations will not be able to capture the complex behaviour of fully-fledged dark energy theories at all times, they should nevertheless recover leading-order effects affecting late universe physics.  
We emphasise that such parametrised and model-independent searches should be seen as an initial coarse tool to identify promising regions of theory space. Specific fundamental theories in these regions can subsequently be further analysed in more targeted searches.

\subsection{Parametrising the background}

In general Horndeski theories, there is sufficient functional freedom such that the Hubble rate $H$ can be set independently of the $\alpha_i$ \cite{Bellini:2014fua}.\footnote{Note that this does not mean that this can be done for any subclass of Horndeski. In quintessence theories, for example, any non-trivial dynamics is associated with a (small) departure from $\Lambda{}$CDM at the background level already. This is somewhat analogous to how slow-roll solutions in inflationary theories are never exactly de Sitter.}  Motivated by the observed proximity of the background expansion to $\Lambda{\rm CDM}$, in what follows we will therefore follow the minimal approach of \cite{BelliniParam,Alonso:2016suf} and fix the background to be that of $\Lambda{\rm CDM}$, considering and constraining perturbations around this background.

The background equations read
\begin{align} \label{back}
H^2 &= \rho_{\rm tot},
&\dot H &= -\frac{3}{2}\left(\rho_{\rm tot}+p_{\rm tot}\right),
\end{align}
where `tot' denotes a sum over all components contributing to the background dynamics (explicitly including the dark energy component) and we note the specific choice of units employed by CLASS and hi\_class, especially $8\pi G = 1$ and a re-scaling of all densities and pressures by a factor of 3.

\subsection{Parametrising linear perturbations: The $\alpha_i$}

Different parametrisations for the $\alpha_i$ are discussed in \cite{Bellini:2014fua,BelliniParam,Linder:2015rcz,Linder:2016wqw,Gleyzes:2017kpi,
Alonso:2016suf,Denissenya:2018mqs,Lombriser:2018olq}. These parametrisations have been used in Refs.~\cite{BelliniParam,Kreisch:2017uet,Alonso:2016suf,Arai:2017hxj,Frusciante:2018jzw,Reischke:2018ooh,Mancini:2018qtb} to both compute and forecast parameter constraints.
However, conclusions about observational constraints on dark energy obtained assuming a specific parametrisation will always be open to the question to what extent that conclusion depends on the specific parametrisation chosen. In order to disentangle physical effects and artefacts of choosing specific parametrisations, we will therefore compute constraints for three different parametrisations (already implemented in hi\_class), which we now summarise:
\\

\nin {\bf Parametrisation I}: A one-parameter ansatz, where the $\alpha_i$ scale with $\Omega_{\rm DE}$
\be \label{oParam}
\alpha_i = c_i \Omega_{\rm DE},
\ee
where we emphasise that $\Omega_{\rm DE}$ here refers to the time-dependent fractional energy density of dark energy, not its value at one specific given time. Linking the parametrisation to $\Omega_{\rm DE}$ ensures that the modification to GR only becomes relevant once dark energy provides a sizeable fraction of the background energy density. This parametrisation is known to accurately capture the evolution of a wide sub-class of Horndeski theories \cite{Pujolas:2011he,Barreira:2014jha}, but not all \cite{Linder:2016wqw}. 
The effective Planck mass $M_S^2$ is inferred from the parametrised $\alpha_M$ via integrating $HM^2_S\hat{\alpha}_M \equiv \frac{d}{dt}M^2_S$.
\\

\nin {\bf Parametrisation II}:  An alternative one-parameter ansatz, with all $\alpha_i$ proportional to the scale factor
\be \label{aParam}
\alpha_i = c_i a.
\ee
The dependence on the scale factor ensures that the modification switches off smoothly at early times (recall that $\alpha_i = 0$ is the GR limit) and is a feature shared by the third parametrisation below as well.
We note that $a$ initially grows more quickly than $\Omega_{\rm DE}$, which only begins to increase at a faster rate than the scale factor around $z=1$, before flattening out eventually. Therefore, dark energy perturbations become relevant slightly earlier in parametrisation II than in parametrisation I. As before, the effective Planck mass $M_S^2$ is inferred from the parametrised $\alpha_M$ via integrating $HM^2_S\hat{\alpha}_M \equiv \frac{d}{dt}M^2_S$.  
\\

\nin {\bf Parametrisation III}: A two-parameter ansatz, where all $\alpha_i$ scale with powers of $a$, except for $\alpha_M$, which is implicitly parametrised via the deviation of the effective Planck mass from (the constant) $M_{\rm Pl}$. Explicitly, we have
\begin{align} \label{apParam}
\alpha_j &= c_j a^{n_j}, &\frac{M_S^2}{M_{\rm Pl}^2} &= 1 + c_{\delta M} a^{n_{\delta M}} \quad \Rightarrow \quad \alpha_M = \frac{c_{\delta M} n_{\delta M} a^{n_{\delta M}}}{M_S^2},
\end{align}
where the index $j$ runs over $\{B,K,T\}$, i.e. braiding, kineticity and gravitational wave speed contributions.
Since the time-dependence of each $\alpha_i$ is freed up individually, different $\alpha_i$ need no longer be proportional to one another here.
A two-parameter ansatz (for each $\alpha_i$) such as \eqref{apParam} has been argued to be well-suited for extracting the maximal information from present data \cite{Gleyzes:2017kpi}. Here we in effect choose the asymptotic late-time value as well as the rate at which the modification switches on, independently for each $\alpha_i$. 
\\

These three parametrisations are suitably rich and different, that any conclusion invariant under a switch between them should be relatively robust and therefore parametrisation-independent. At the same time, considering different parametrisations will also allow us to get an understanding of which features are a consequence of choosing a specific parametrisation, rather than a conclusion enforced by the data themselves.

\subsection{Stability conditions}

Imposing stability conditions on the parameters of the theory serves two purposes. Firstly, observationally relevant instabilities would exclude the associated parameter values in any case, so checking for their potential presence before computing the full cosmology in an MCMC run increases computational efficiency, but does not alter the result. Gradient instabilities are frequently of this type. Secondly, some instabilities may not show up in the classical analysis one performs in an MCMC run, but nevertheless undermine the validity of the theory (e.g. once quantum effects are taken into account). Ghost instabilities can be of this type, leading to an exponential decay of the vacuum that any purely classical analysis would be blind to. Checking whether ghost instabilities are present therefore safeguards against accidentally including theories that are ill-defined at a fundamental level. 

We will impose the standard stability conditions implemented by hi\_class. These are firstly ghost-freedom conditions for the scalar and tensor mode, respectively given by
\begin{align}
\alpha_K + \tfrac{3}{2}\alpha_B^2 &> 0, &M_S^2 &> 0.
\end{align}
If these were broken in the linear theory already, this is the sign of a fatal instability for the theory. Note that for Horndeski theories the no-ghost condition is explicitly k-independent, so a would-be ghost is present at all scales/energies here. In more general modified gravity theories k-dependence can enter into a no-ghost condition \cite{Lagos:2017hdr}, in which case a more careful analysis of the precise nature of the ghost is required (e.g. small $k$ ghosts have been argued to be harmless in \cite{Gumrukcuoglu:2016jbh}).

Secondly, we impose the absence of gradient instabilities, i.e. a positive speed of sound (effectively this amounts to considering the large-$k$ limit of the `mass' term in Fourier space and requiring this term to be positive). In Horndeski theories with $\alpha_T = 0$, i.e. for \eqref{Horndeski_simple}, the condition for the absence of a tensor mode gradient instability is trivially satisfied, while the scalar mode condition is given by 
\be \label{gradient_condition_1}
c_s^2 = \frac{1}{\alpha_K + \tfrac{3}{2}\alpha_B^2} \cdot \left[ (2-\alpha_B)\left(\tfrac{1}{2}\alpha_B + \alpha_M - \frac{\dot H}{H^2}\right) - \frac{3(\rho_{\rm tot} + p_{\rm tot})}{H^2M_S^2} + \frac{\dot \alpha_B}{H}\right] > 0.
\ee
Using the background equations \eqref{back}, we will find it useful to recast this condition in the equivalent form
\be \label{gradient_condition_2}
c_s^2 = \frac{1}{\alpha_K + \tfrac{3}{2}\alpha_B^2} \cdot \left[ (2-\alpha_B)\left(\tfrac{1}{2}\alpha_B + \alpha_M\right) +\frac{2 \dot H}{H^2}\left(\frac{1-M_S^2}{M_S^2}\right) + \frac{\tfrac{d}{dt}{(\alpha_B H)}}{H^2}\right] > 0. 
\ee
These conditions need to be modified, if the speed of gravitational waves is left undetermined. In this case, the absence of gradient instabilities for the tensor mode requires
\be
\alpha_T \geq -1,
\ee
while for the scalar mode the gradient stability condition becomes
\be \label{gradient_condition_3}
c_s^2 = \frac{1}{\alpha_K + \tfrac{3}{2}\alpha_B^2} \cdot \left[ (2-\alpha_B)\left(\tfrac{1}{2}\alpha_B(1+\alpha_T) + \alpha_M - \alpha_T - \frac{\dot H}{H^2}\right) - \frac{3(\rho_{\rm tot} + p_{\rm tot})}{H^2M_S^2} + \frac{\dot \alpha_B}{H}\right] > 0.
\ee
We emphasise that using the above gradient stability conditions to exclude parameter space in exploring theories ought to be used with caution. Cosmologies with significant such classical instabilities will automatically be excluded by the data, when exploring the full parameter space with an MCMC run. While it is computationally more efficient to only explore a reduced parameter space based on the above stability cuts, one ought to be careful not to place overzealous cuts and exclude physically viable parameter space (thus biasing the results). We have therefore checked for a number of cases that constraints with and without gradient stability priors indeed only show very mild differences, so using the above cuts is well-justified.
Tachyon instabilities and the associated stability conditions, on the other hand, are far more involved (for a discussion see \cite{Lagos:2017hdr,DeFelice:2016ucp,Frusciante:2018vht}), especially since the presence of such instabilities can in fact be required to ensure the physical validity of a model (the Jeans instability is the prime example here). We therefore choose a maximally safe approach and do not exclude any parts of parameter space based on tachyonic stability cuts a priori, but instead let the data exclude any cosmologies with significant such instabilities. While this approach is less computationally efficient, it guards against biasing our results by only sampling part of the physically viable parameter space. 
\\

\section{Data and theoretical modelling}\label{sec:data}

In order to constrain general Horndeski scalar-tensor theories, we combine several different data sets, which are illustrated in Tab.~\ref{tab:data_sets} and detailed below.\footnote{We note that there exist many data sets additional to the ones discussed below. In this work, we choose a conservative approach and exclude any data sets that could be correlated to each other, as they probe the same underlying structures.}

\subsection{Data}

\nin \textbf{CMB:} We use CMB temperature and polarisation data from Planck's second data release \cite{Planck-Collaboration:2016ad, Planck-Collaboration:2016af, Planck-Collaboration:2016ae}. Specifically, we include low-$\ell$ ($2 \leq \ell \leq 29$) temperature and polarisation data as well as high-$\ell$ ($30 \leq \ell \leq 2508$) temperature data in form of the published Planck likelihood. In our fiducial setup, we analyse high-$\ell$ temperature data using the \texttt{Plik lite} likelihood, which has been marginalised over all nuisance parameters except for the Planck absolute calibration parameter.
\footnote{We have tested the impact of this choice, by re-running our analysis for a fiducial $\Lambda{}$CDM background cosmology using the full Planck high-$\ell$ temperature likelihood (with all of its additional nuisance parameters) instead of the pre-marginalised \texttt{Plik lite} likelihood. Constraints obtained using the full Planck high-$\ell$ temperature likelihood and ones obtained using the \texttt{Plik lite} likelihood agree very well -- see appendix \ref{ap:extra} for details. Note that this was to be expected, due to the explicit choice of a $\Lambda{}$CDM background cosmology in our analysis -- see the discussion in section \ref{sec:param}.}. 
We further complement these measurements with the Planck CMB lensing likelihood \cite{Planck-Collaboration:2016aa} in the angular multipole range $40 \leq \ell \leq 400$.
\\

\nin \textbf{RSD:} We include two complementary RSD measurements in our analysis. As a first data set, we use the RSD measurement derived from BOSS DR11 CMASS anisotropic clustering at an effective redshift $z_{\mathrm{eff}} = 0.57$ \cite{Samushia:2014}. Further, we also include the RSD measurement at $z_{\mathrm{eff}} = 0.067$ obtained from 6dF galaxy clustering data in Ref.~\cite{Beutler:2012}.
\\

\nin \textbf{BAO:} We complement the above data sets with isotropic BAO measurements from BOSS and SDSS. Specifically we include constraints on the volume averaged distance $D_{V}$ at $z_{\mathrm{eff}} = 0.32$ from BOSS DR11 LOWZ data \cite{Anderson:2014} and BAO measurements at $z_{\mathrm{eff}} = 0.15$ from the SDSS DR7 main sample \cite{Ross:2015}. Note that we exclude the anisotropic BAO measurement from CMASS, which is also given in Ref.~\cite{Anderson:2014}, as it is highly correlated with the RSD measurement of Ref.~\cite{Samushia:2014}. Ref.~\cite{Ross:2015} showed that the cross-correlations between the RSD and BAO data sets included in this work are negligible and we thus assume all these data sets to be independent.
\\

\nin \textbf{mPk:} Finally we include constraints on the shape of the matter power spectrum (mPk) at $z_{\mathrm{eff}} = 0.35$ from SDSS DR4 luminous red galaxies (LRG) \cite{Tegmark:2006}. Ref.~\cite{Tegmark:2006} measured the galaxy clustering power spectrum $P_{gg}(k)$ in three-dimensional Fourier space for 20 $k$-bands in the range $0.01 h\mathrm{Mpc}^{-1} < k < 0.2 h\mathrm{Mpc}^{-1}$. In our analysis, we only consider $k$-bands with $k < 0.1 h\mathrm{Mpc}^{-1}$ in order to minimise sensitivity to non-linear clustering and scale-dependent bias. 

\begin{table*}
\caption{Overview of the data sets considered in this work.} 
\begin{center}
\begin{tabular}{>{\centering}m{2.6cm}>{\centering}m{8cm}>{\centering}m{2.5cm}@{}m{0pt}@{}} \hline\hline 
Acronym & Description & Reference & \\ \hline \Tstrut     
P15 & Constraints from Planck Collaboration 2015, TT+lowP & \cite{Planck-Collaboration:2016ae} & \\ \\
P15+lensing & Constraints from Planck Collaboration 2015, TT+lowP and CMB lensing convergence. & \cite{Planck-Collaboration:2016ae, Planck-Collaboration:2016aa} & \\ \\
BAO & BAO measurements from BOSS and SDSS. \\ 
	    SDSS DR7: $z_{\mathrm{eff}} = 0.15$, \\
	    $D_{V}(z_{\mathrm{eff}}) = (664 \pm 25) \sfrac{r_{d}}{r_{d, \mathrm{fid}}}$ Mpc \\
	    BOSS DR11:  $z_{\mathrm{eff}} = 0.32$, \\
	    $D_{V}(z_{\mathrm{eff}}) = (1264 \pm 25) \sfrac{r_{d}}{r_{d, \mathrm{fid}}}$ Mpc& \cite{Anderson:2014, Ross:2015} & \\ \\
RSD & RSD constraints from BOSS and 6dF. 
	    \\ BOSS DR11: $z_{\mathrm{eff}} = 0.57$, \\
	    $\sfrac{D_{V}(z_{\mathrm{eff}})}{r_{d}} = 13.85 \pm 0.17$, \\
	    $F(z_{\mathrm{eff}}) = 0.6725 \pm 0.0283$, \\
	    $f(z_{\mathrm{eff}})\sigma_{8}(z_{\mathrm{eff}}) = 0.4412 \pm 0.0435$ 
	    \\ 6dF: $z_{\mathrm{eff}} = 0.067$, \\
	    $f(z_{\mathrm{eff}})\sigma_{8}(z_{\mathrm{eff}}) = 0.423 \pm 0.055$& \cite{Beutler:2012, Samushia:2014} & \\ \\
mPk & Constraints from SDSS DR4 LRG power spectrum shape & \cite{Tegmark:2006} & \\ \hline \hline
\end{tabular}
\end{center}
\label{tab:data_sets}
\end{table*} 

\subsection{Theoretical modelling}

We compute theoretical predictions for all observables considered using the publicly-available code \texttt{hi\_class}\footnote{The code can be found at: \texttt{http://miguelzuma.github.io/hi\_class\_public/}.}\cite{hiclass}, which extends the Boltzmann code \texttt{class}\footnote{The code can be found at: \texttt{http://class-code.net}.} \cite{Blas:2011rf} to subsets of Horndeski scalar-tensor theory \cite{Horndeski:1974wa}. For CMB and BAO data we follow the implementations described in Refs.~\cite{Planck-Collaboration:2016af, Planck-Collaboration:2016aa, Anderson:2014, Samushia:2014}. Detailed explanations of theoretical modelling choices employed for RSD and matter power spectrum data are given in Appendix \ref{ap:modeling_details}.

\section{Cosmological parameter constraints} \label{sec:constraints}

We derive constraints on cosmological parameters in a joint fit to the data discussed in Sec.~\ref{sec:data}. We make the simplifying assumption that the cross-correlations between all data sets are negligible, and we therefore assume a joint Gaussian likelihood as
\begin{equation}
\mathscr{L}(D | \boldsymbol{\theta}) = \mathscr{L}_{\mathrm{CMB}}(D_{\mathrm{CMB}} | \boldsymbol{\theta}) \mathscr{L}_{\mathrm{RSD}}(D_{\mathrm{RSD}} | \boldsymbol{\theta}) \mathscr{L}_{\mathrm{BAO}}(D_{\mathrm{BAO}} | \boldsymbol{\theta}) \mathscr{L}_{\mathrm{mPk}}(D_{\mathrm{mPk}} | \boldsymbol{\theta}),
\end{equation}  
where $\boldsymbol{\theta}$ denotes the vector of model parameters and $D_{i}$ a given data vector.
We sample $\mathscr{L}(D | \boldsymbol{\theta})$ in a Monte Carlo Markov Chain (MCMC) with the publicly-available code \texttt{MontePython}\footnote{The code can be found at: \texttt{http://baudren.github.io/montepython.html}.} \cite{Audren:2012wb,Brinckmann:2018cvx}, using the Metropolis-Hastings algorithm \cite{Metropolis:1953, Hastings:1970}. We set the background cosmological model to $\Lambda$CDM and, in addition to the parameterisation-dependent modified gravity parameters detailed in section \ref{sec:param}, vary the six cosmological parameters $\{w_{\mathrm{cdm}}, \, \allowbreak w_{\mathrm{b}}, \, \allowbreak \theta_{\mathrm{s}}, \allowbreak  n_{\mathrm{s}}, \,\allowbreak \log{10^{10}A_{s}}, \, \allowbreak \tau_{\mathrm{reion}}\}$, where $w_{\mathrm{cdm}} = \Omega_{\mathrm{cdm}}h^{2}$ is the fractional cold dark matter density today, $h$ is the Hubble parameter, $w_{\mathrm{b}} = \Omega_{\mathrm{b}}h^{2}$ is the fractional baryon density today, $\theta_{\mathrm{s}}$ is the position of the first peak in the CMB temperature anisotropy power spectrum, $n_{\mathrm{s}}$ denotes the scalar spectral index of initial perturbations, $A_{\mathrm{s}}$ is the primordial power spectrum amplitude at a pivot scale of $k_{0} = 0.05$ Mpc$^{-1}$ and $\tau_{\mathrm{reion}}$ denotes the optical depth to reionisation. We also suppress the tensor-to-scalar ratio, setting $r=0$, and impose that the asymptotic value of the effective Planck mass $M$ at early times is indeed $M_{\rm Pl}$, since we do not wish to constrain early universe modifications of gravity (for a different approach, see \cite{BelliniParam}). Following Ref.~\cite{Planck-Collaboration:2016ae}, our fiducial model includes two massless and a massive neutrino eigenstate and we fix the sum of their masses to the minimal mass allowed by oscillation experiments, i.e. $\sum_{\nu} m_{\nu} = 0.06$ eV. In addition to the cosmological parameters, we further vary three nuisance parameters $\{A_{\mathrm{Planck}} \, \allowbreak, b, \, \allowbreak n\}$, where $A_{\mathrm{Planck}}$ denotes the Planck absolute calibration parameter, $b$ is a linear, redshift-independent galaxy bias parameter and $n$ parametrises systematic uncertainties due to shot noise and nonlinear evolution in the matter power spectrum (for more details, see Appendix \ref{ap:modeling_details}). When extracting parameter constraints we check for convergence, in particular ensuring the Gelman-Rubin diagnostic \cite{Gelman:1992zz} $R$ satisfies $R - 1 \lesssim 0.01$ for all (cosmological and nuisance) parameters.

\subsection{Constraining the $\alpha_i$}

\begin{figure}%
    \centering
    \subfloat[$\alpha_i = c_i \Omega_{\rm DE}$]{{\includegraphics[width=0.49\linewidth]{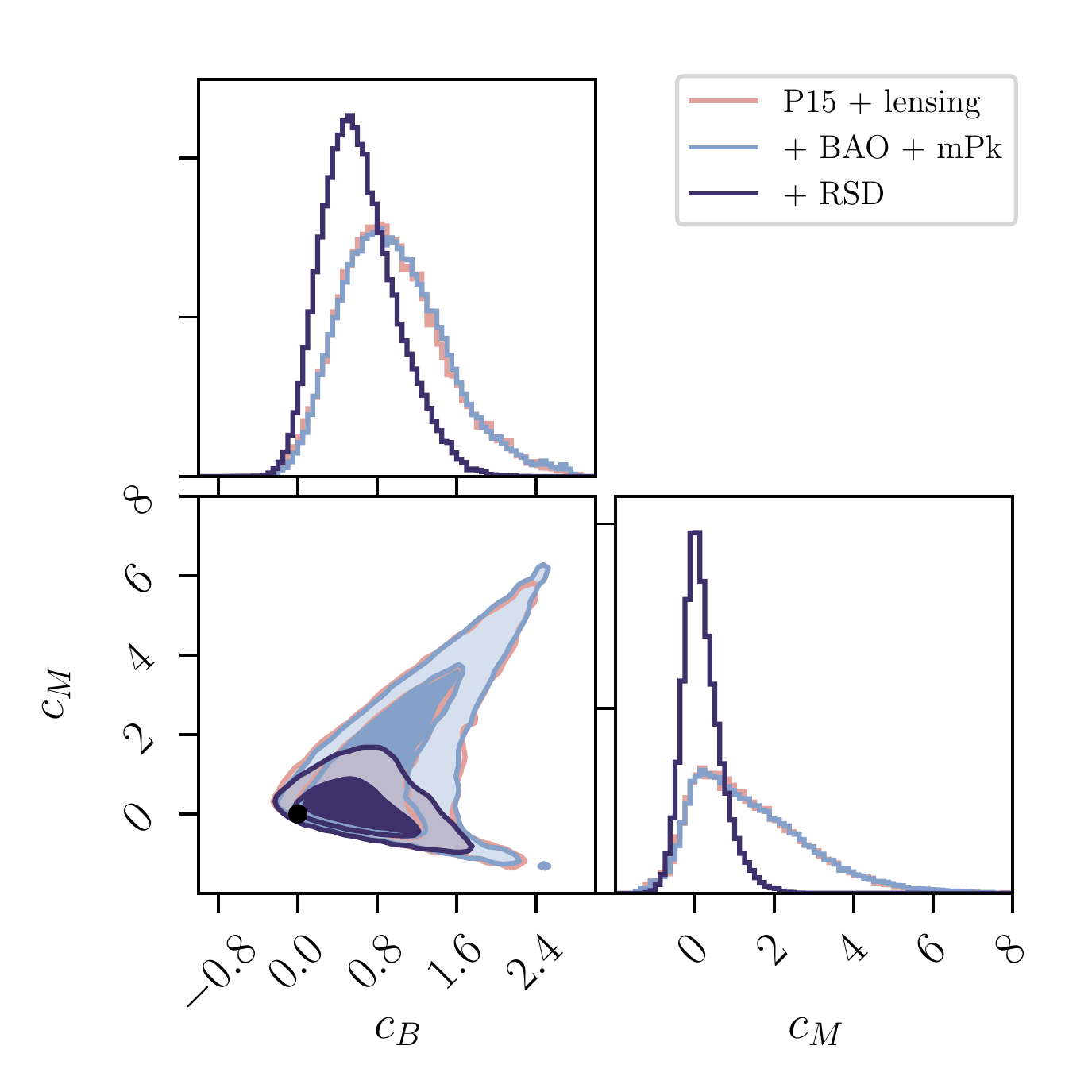} }}%
    \subfloat[$\alpha_i = c_i a$]{{\includegraphics[width=0.49\linewidth]{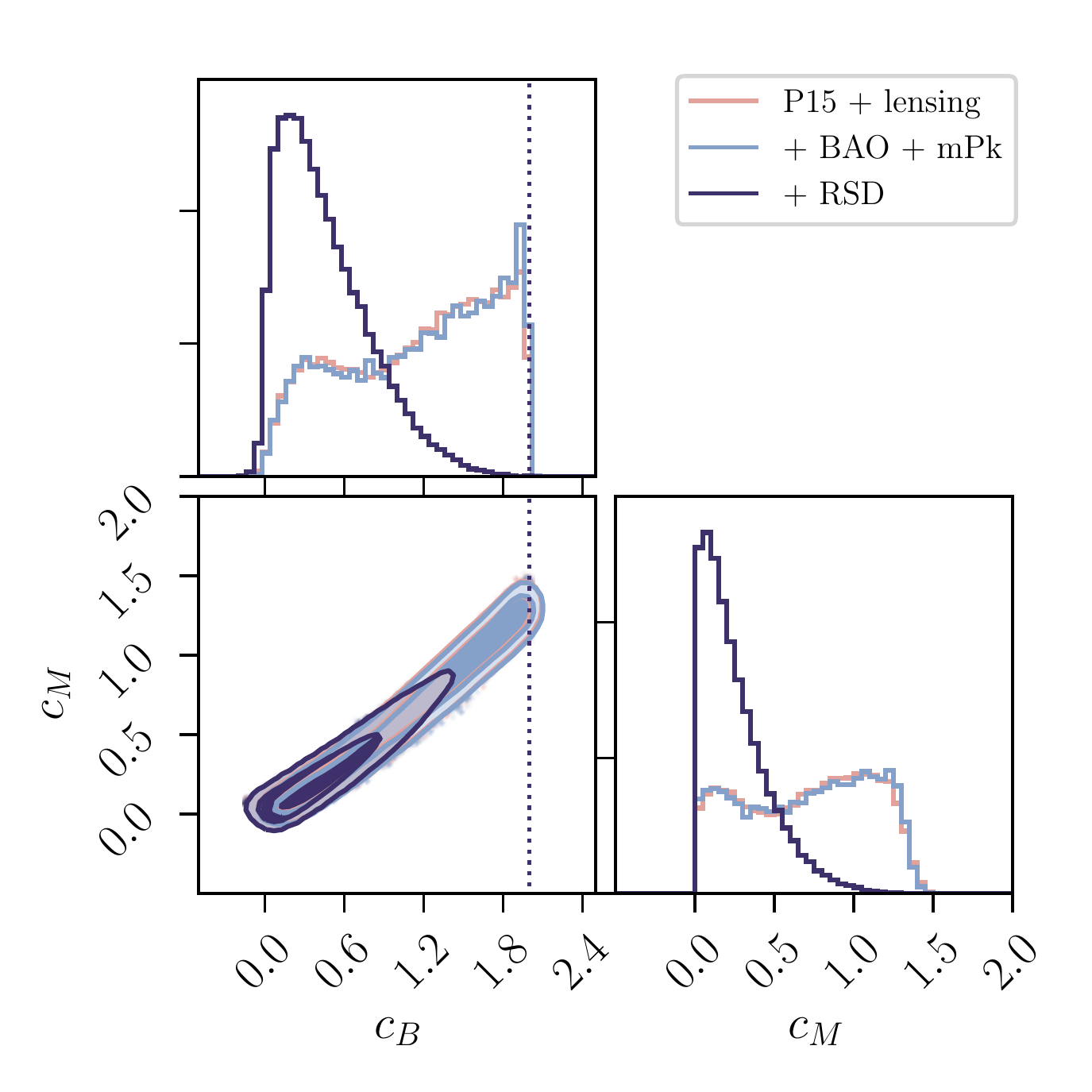} }}%
    \caption{Cosmological parameter constraints for the modified gravity $c_i$ parameters, using parametrisations \eqref{oParam} and \eqref{aParam}. The inner (outer) contours correspond to 68$\%$ (95$\%$) confidence levels, respectively and we plot results for different combinations of the datasets detailed in section \ref{sec:data} above. The lower (negative) $c_M$ boundary is due to the onset of gradient instabilities. Otherwise, without RSD data, the shape of the contours is primarily driven by the late ISW effect in the low-$\ell$ CMB temperature anisotropy power spectrum $C_\ell^{TT}$ (also see figure \ref{fig:Cl_plot}). Once RSDs are taken into account, their measurement of $f\sigma_8$ establishes a strong upper bound for $c_M$, thus strengthening constraints. Finally note one additional feature for the \eqref{aParam} parametrisation (right panel). All models with $c_B > 2$ here will cross the singular $\alpha_B = 2$ point in their past evolution (for the left panel this would correspond to $c_B \sim 2.86$, so does not affect constraints there) and we consequently do not explore such models, as discussed below.}%
    \label{fig:aoParam}%
\end{figure}

\begin{figure}[t]
\begin{center}
\includegraphics[width=0.7\linewidth]{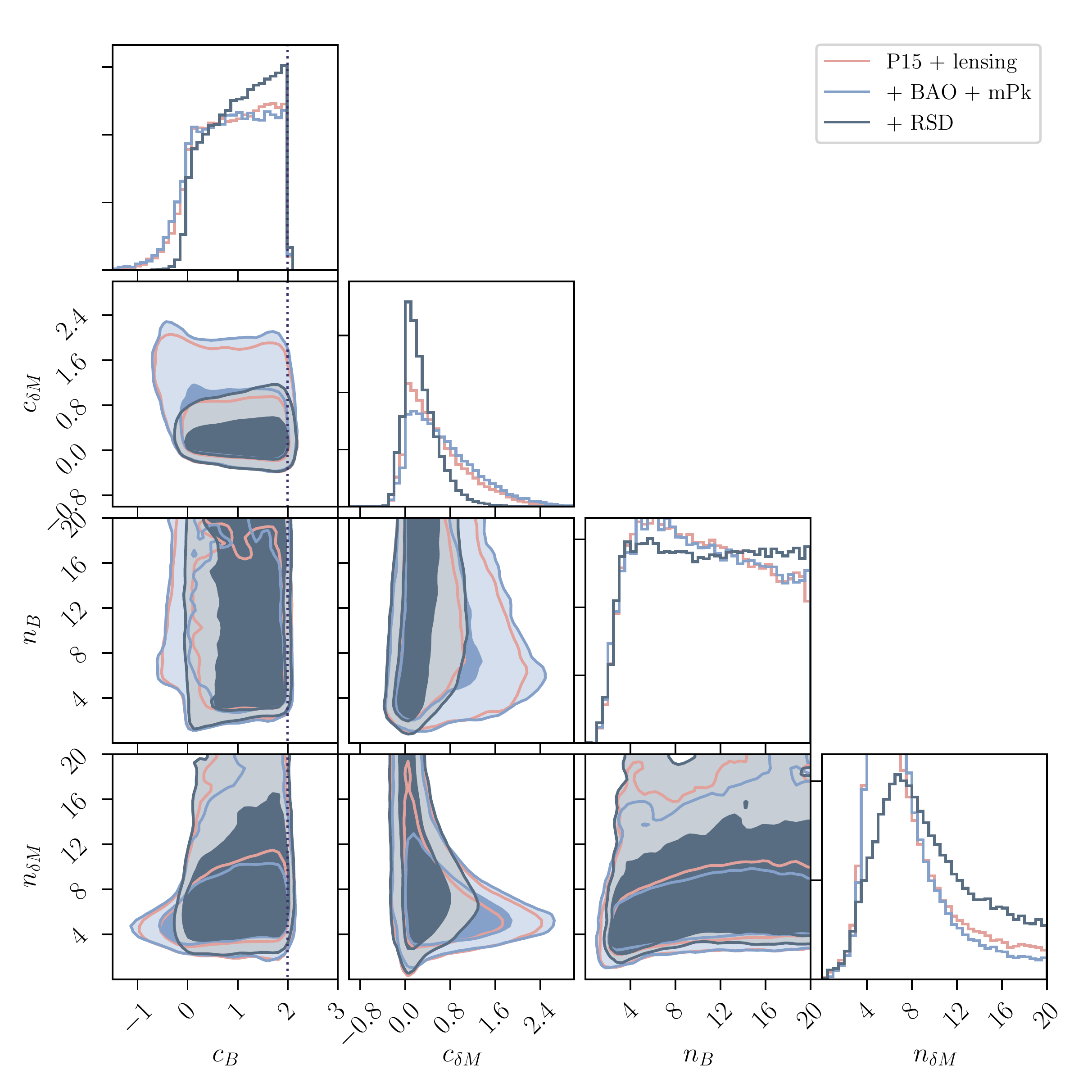}
\end{center}
\caption{Cosmological parameter constraints for the modified gravity parameters in the \eqref{apParam} parametrisation.
The inner (outer) contours correspond to 68$\%$ (95$\%$) confidence levels, respectively and we plot results for the same datasets as in figure \ref{fig:aoParam}. In close analogy to that figure, $c_M$ is tightly constrained, whereas freeing up the power-law dependence on $c_B$ weakens the constraints for that parameter, illustrating that the $c_B$ constraints in figure \ref{fig:aoParam} where significantly strengthened by choosing a parametrisation for which $\alpha_B \propto \alpha_M$. Also again notice the singularity cut for $c_B = 2$. We impose bounds for the poorly constrained power-law parameters $n_i \leq 20$, since the analysis will not converge otherwise: Arbitrarily large $n_i$ correspond to suppressing modified gravity effects until extremely late in the evolution, so very large $n_i$ yield near-identical cosmological phenomenology. The constraint for $n_{\delta M}$ is stronger than that for $n_B$, since in this parametrisation $c_{\delta M} n_{\delta M}$ is analogous to $c_M$ in the above parametrisations -- recall that here $\alpha_j = c_j a^{n_j}$ and  $M_S^2 = M_{\rm Pl}^2(1 + c_{\delta M} a^{n_{\delta M}})$ so $M_S^2 \alpha_M = c_{\delta M} n_{\delta M} a^{n_{\delta M}}$. 
\label{fig:apParam}}
\end{figure}

\begin{table*}
\begin{center}
\begin{tabular}{>{\centering}m{2.5cm}cc} \hline\hline 
Parametrisation & Parameter & Posterior\\ \hline \Tstrut                             
\multirow{2}{*}{I} & $c_{B}$ & $0.63\substack{+0.83 \\ -0.62}$ \\ 
 & $c_{M}$ & $0.20\substack{+1.15 \\ -0.82}$ \\ \hdashline \Tstrut
\multirow{2}{*}{II} & $c_{B}$ & $0.48\substack{+0.83 \\ -0.46}$ \\ 
 & $c_{M}$ & $0.27\substack{+0.54 \\ -0.26}$ \\ \hdashline \Tstrut
\multirow{2}{*}{III} & $c_{B}$ & $1.1\substack{+0.89 \\ -1.1}$ \\
& $c_{\delta M}$ & $0.30\substack{+0.77 \\ -0.45}$ \\ 
  \hdashline \Tstrut
\multirow{3}{*}{\parbox{2.5cm}{\centering I ($c_{T}$ free)}} & $c_{B}$ & $0.71\substack{+0.88 \\ -0.71}$ \\ 
 & $c_{M}$ & $-0.01\substack{+1.3 \\ -0.87}$ \\ 
 & $c_{T}$ & $<0.26$ \\ 
 \hline\hline 
\end{tabular}
\end{center}
\caption{Constraints on the modified gravity parameters for the different parametrisations used in this work. Note that we do not include the $n_i$ parameters of parametrisation III \eqref{apParam} here, since these are only very poorly constrained -- see figure \ref{fig:apParam}. The uncertainties/limits quoted denote the $95 \%$ c.l.. In the final parametrisation, $c_T$ has a highly skewed, non-Gaussian posterior (see figure \ref{fig:cT}), so we only give an upper limit at the $95 \%$ c.l. for this parameter.}
\label{tab:params}
\end{table*}

In the following, we show parameter constraints for the $\hat\alpha_i$ functions that parametrise departures from GR for various combinations of the datasets listed in table \ref{tab:data_sets} and for the different $\hat\alpha_i$-parametrisations, \eqref{oParam}, \eqref{aParam} and \eqref{apParam}. In this section we assume luminally propagating gravitational waves, so work with \eqref{Horndeski_simple} as the underlying action. Theories with $c_{\rm GW} \neq c$ will be discussed in section \ref{sec:ct}. Constraints shown are marginalised over all standard $\Lambda{}$CDM and nuisance parameters as discussed above. The reader is referred to appendix \ref{ap:extra} for further details on parameter constraints for these standard parameters and consistency checks. 
\\

\nin{\bf The constrainable $\hat\alpha_i$ parameter space}: Out of the four $\alpha_i$, it is important to note that $\alpha_K$ is in effect the combination of the $G_i$ and their derivatives that is 'orthogonal' to the parameter space probed by linear cosmology and therefore hardly constrained by the data used here~\cite{BelliniParam}. 
For the purposes of the analysis here, one can consequently fix $\alpha_K$ to an essentially arbitrary fiducial parameter, and we will do so in what follows.\footnote{We have checked that our results do not change for a wide variety of fiducial choices for $\alpha_K$. For concreteness we have chosen $c_K = 0.1$ for all the results shown here. $n_K$, in the context of parametrisation III, has been fixed to $n_K = 3$.} 
In addition, we highlight that $\alpha_B = 2$ is a singular point in the $\alpha_i$ parameter space\footnote{We thank Emilio Bellini for several discussions related to this point.} in the following sense: When computing the linear theory around an FRW background in the context of Horndeski models \eqref{Horndeski_action}, one may focus on the scalar perturbations of the metric and the scalar $\phi$, expand the action to quadratic order in these perturbations, gauge fix and integrate out auxiliary variables (for details, see \cite{Lagos:2017hdr}). This procedure results in a kinetic term for the scalar perturbation $\delta\phi$ of the following form
\begin{align}
{\cal L}^{(2)}_{\rm kin} &\propto \frac{\left(3 \hat{\alpha}_\textrm{B}^2 + 2 \hat{\alpha}_\textrm{K}\right) H^2 M_S^2 \dot{\delta\phi^2}}{\left(\hat{\alpha}_\textrm{B} - 2\right)^2}.\label{STkin}
\end{align}
Clearly the kinetic term diverges when $\hat{\alpha}_B=2$. This is because setting $\hat{\alpha}_B=2$ eliminates mixing between the metric perturbations $\Phi$ and $B$ in the action, turning $B$ into a Lagrange multiplier -- for details we again refer to \cite{Lagos:2017hdr}. As a result, the $\hat{\alpha}_B=2$ theory propagates no gravitational scalar degree of freedom and no physical model should therefore evolve across this boundary, as this would imply a discontinuity in the number of propagating \dofs{}. In the parameter plots we are about to discuss, we will therefore explicitly mark the $\hat\alpha_B = 2$ line whenever relevant and forbid evolution across this boundary.\footnote{Note that the plots in figures \ref{fig:aoParam} and \ref{fig:apParam} present binned data, so any points seemingly just over the $\hat\alpha_B = 2$ boundary appear as such as an artefact of the binning.}
\\
 
\nin {\bf Parametrisation I}: $\alpha_i = c_i \Omega_{\rm DE}$. Constraints for this parametrisation are shown in figure \ref{fig:aoParam}. First, we note that the sharp, lower (small $c_M$) boundary of the contours is due to the onset of gradient instabilities. For the CMB-only constraints, the other boundaries are mostly determined by the late integrated Sachs-Wolfe (ISW) effect. The ISW especially excludes cosmologies with large $c_M$ or $c_B$, as these lead to the generation of too much power for the low-$\ell$ $C_{\ell}^{TT}$, which can be seen in figure \ref{fig:Cl_plot}. This is in good agreement with the observation of \cite{hiclass} that modifications to the late ISW are indeed the driving factor in CMB constraints on the $c_i$ -- also see \cite{Kreisch:2017uet}. Note that from CMB constraints alone there is a `degeneracy direction' roughly satisfying $c_M \sim 2.5 c_B$ for small $c_M$, where the effect of the $c_i$ conspires to avoid large power for the low-$\ell$ $C_{\ell}^{TT}$. Adding BAO and mPk data only very mildly improves constraints, whereas RSD constraints on $f\sigma_8$ rule out large, positive $c_M$ values.
Finally note that the singular $\alpha_B = 2$ case discussed above does not appear here, as in this parametrisation and for $\Omega_{DE, 0} \sim 0.7$, $\alpha_{B} = 2$ today corresponds to $c_B \sim 2.86$ (or progressively larger as one goes back in time) and these high-$c_B$ cosmologies are already strongly disfavoured by Planck CMB constraints.  
\\

\nin {\bf Parametrisation II}: $\alpha_i = c_i a$. As can be seen from Fig.~\ref{fig:aoParam}, analogously to above, the CMB-only constraints lie along a `degeneracy direction',  which in this parametrisation is given by $c_b \sim 1.8 c_M$. Also as before, BAO and mPk data do not add significant additional constraining power. The contours around the `degeneracy direction' are tightened in comparison with the $\alpha_i = c_i \Omega_{DE}$ parametrisation discussed above, resulting in a correspondingly tighter correlation between $c_M$ and $c_B$ -- departures away from this direction lead to large excess power on large scales again.
Also in contrast to the first parametrisation, all negative $c_M$ values are ruled out by requiring the absence of gradient instabilities in this parametrisation. Finally notice that constraints not including RSDs in fact favour large values of $c_B$ and $c_M$, driving chains in the analysis to preferentially explore regions close to the singular point $\alpha_B = 2$ (which is crossed in the past for $c_B > 2$). Combined with a somewhat bi-modal distribution for both $c_i$ when only using these datasets, this leads to slowly converging chains for these cases. However, RSDs rule out large $c_B,c_M$ values, thus driving the preferred values for the $c_i$ back closer to GR and removing any bi-modality.
\\

\nin {\bf Parametrisation III}: $\alpha_j = c_j a^{n_j}$.
The constraints for parametrisation III are shown in figure \ref{fig:apParam}. In order to correctly interpret the results, recall that $M_S^2 \alpha_M = c_{\delta M} n_{\delta M} a^{n_{\delta M}}$ here, so the analogue to $c_M$ in the previous parametrisations is $c_{\delta M} n_{\delta M}$, not $c_{\delta M}$. 
We first note that we impose an upper bound on the $n_i$, namely $n_i \leq 20$, as large $n_i$ essentially remove any observable cosmological effect of the $\alpha_i$ by suppressing them until very late times.\footnote{Negative $n_i$ introduce large modifications at early times, so we do not discuss this case, since we are focusing on late-time modifications here. The precise upper bound on $n_i$ is arbitrary, but we have chosen $n_i \leq 20$ to facilitate the comparison with \cite{Kreisch:2017uet}.} Large $n_i$ cosmologies are therefore indistinguishable from standard $\Lambda{}$CDM in practice, so that MCMCs will not converge unless an upper bound is placed on the $n_i$.
This is particularly manifest in the constraints for $n_B$. $n_{\delta M}$ is far better constrained, precisely for the above reason that $c_M$ in the above parametrisations is analogous to $c_{\delta M} n_{\delta M}$ here, so $n_{\delta M}$ inherits some of the constraining power acting on $c_M$. In addition, removing the proportionality between $\alpha_M$ and $\alpha_B$ removes any significant correlation between $c_{\delta M}$ and $c_B$. This can be seen by noting that $c_{\delta M}$ is still driven to low values by RSD constraints, as before, but this does not lead to a corresponding tightening of constraints on $c_B$ here. $c_B$ is only weakly affected by adding RSD constraints and for all datasets prefers values close to the singular $c_B = 2$ point. 
Note that parameter constraints using this parametrisation have also been derived and discussed in \cite{Kreisch:2017uet}. Our analysis differs from the results presented there in two important aspects.
First, we take into account additional modified gravity effects on $f\sigma_8$ in our analysis (see appendix \ref{ap:modeling_details}), resulting in stronger constraints from RSDs.
Secondly, as discussed in section \ref{sec:param}, we do not exclude models that display tachyonic instabilities. Since such instabilities can be an essential part of well-motivated physical models, we simply let the data decide which models to accept. In this way one is sure to avoid introducing unphysical biases as artefacts of overzealous stability priors.
Overall the differences in the analysis have a strong effect on the parameter constraints obtained: \cite{Kreisch:2017uet} found that the posterior for $c_{\delta M}$ (or $\tilde M_0$ in the notation used in \cite{Kreisch:2017uet}) displayed strong bi-modality and was predominantly driven away from its GR value zero, as a result of tachyonic instabilities.\footnote{We thank Christina Kreisch for related discussions.}
Our analysis shows no such bimodality for the $c_{\delta M}$ posterior, qualitatively changing constraints on the running of the Planck mass in comparison to \cite{Kreisch:2017uet}, with the best-fitting cosmologies clustered around the GR value for this parameter.
As can be seen from figure \ref{fig:apParam}, this change in posterior for $c_{\delta M}$ also qualitatively changes the constraints for the other parameters, removing any strong suppression for large values of $n_{\delta M}$ in the associated posterior (and similarly removing any suppression for large values of $n_B$ in its posterior). 
\\

\subsection{What drives the constraints?} \label{subsec:drive}

\begin{figure}[t]
\begin{center}
\includegraphics[width=0.7\textwidth]{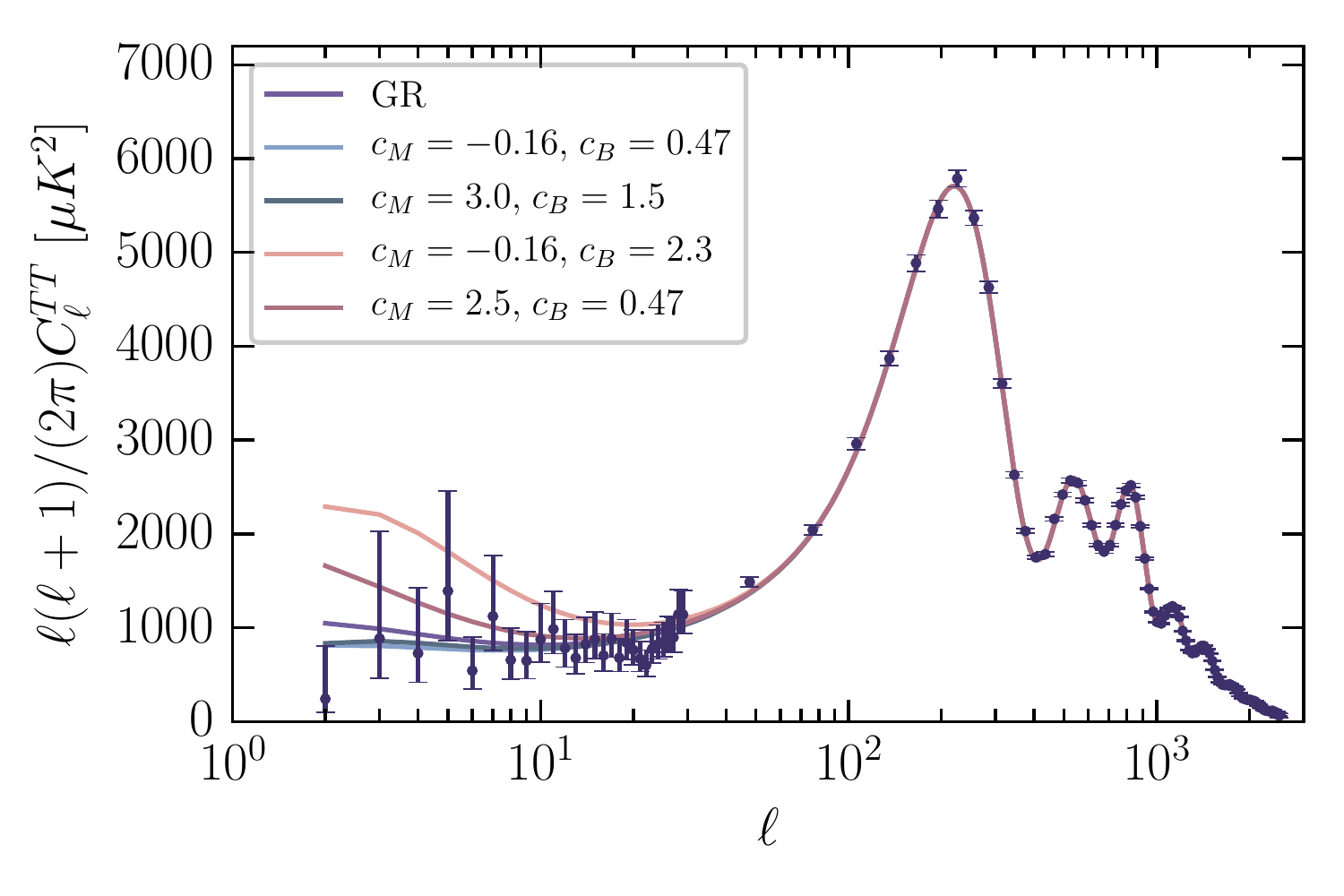}
\end{center}
\caption[CMB TT power spectrum]{Illustration of the effect of modified gravity parameters on the CMB TT power spectrum. The data points used in this work are shown with $1\sigma$ uncertainties\footnotemark and we plot a standard $\Lambda{}$CDM/GR cosmology as well as four other cosmologies with non-vanishing $c_i$. Since we focus on the effects of the $c_i$, all standard $\Lambda{}$CDM parameters are fixed to their Planck 2015 best-fit values here \cite{Planck-Collaboration:2016ae}. Note that the second ($c_M = -0.16, c_B = 0.47$) cosmology corresponds to the best-fit values we obtain for these parameters in our MCMC analysis for the $\alpha_i = c_i \Omega_{DE}$ parametrisation (although the corresponding cosmology here is close to, but not identical, to that best-fit, since we impose the Planck best-fit choices for all other parameters). The third cosmology shows that there is a `degeneracy' direction associated with simultaneously enlarging $c_M$ and $c_B$ from CMB constraints alone (and only for `small' $c_i$, see figure \ref{fig:aoParam}). The final two cosmologies illustrate that individually increasing $c_M$ or $c_B$ eventually leads to too much added power on large scales via the late ISW effect.
\label{fig:Cl_plot}}
\end{figure}

\begin{figure}
\begin{center}
\includegraphics[width=0.99\textwidth]{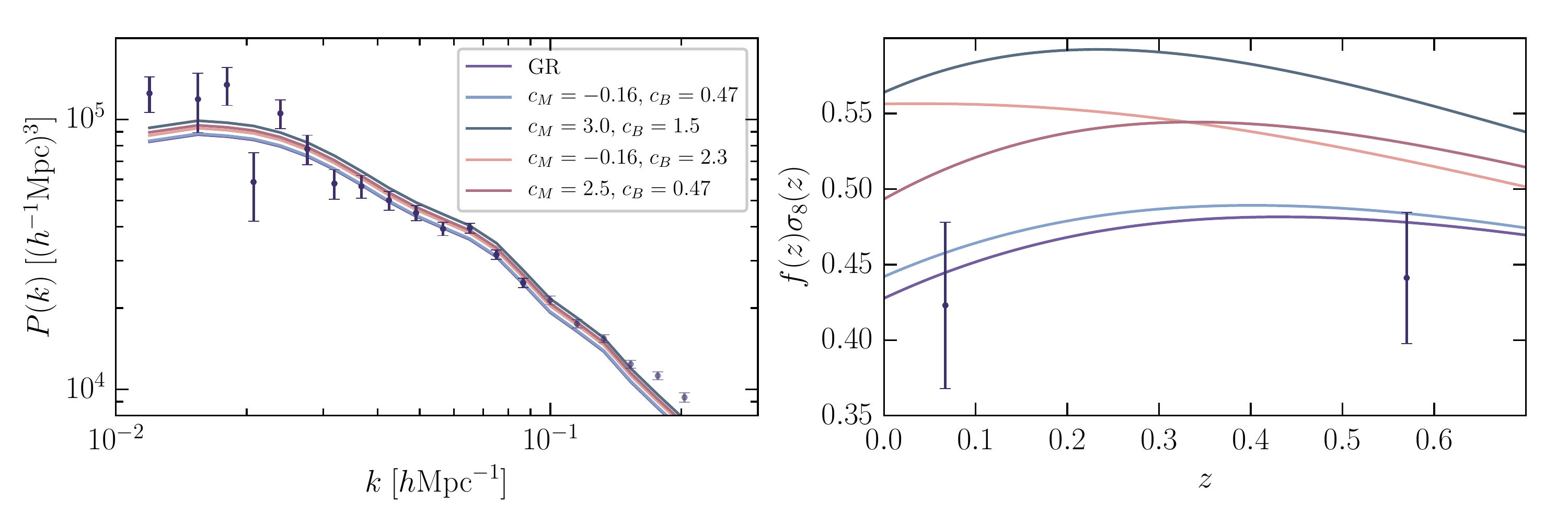}
\caption{Illustration of the effect of modified gravity parameters on the matter power spectrum and $f\sigma_8$. The data points used in this work are shown with $1\sigma$ uncertainties as before and we plot the same set of cosmologies as in figure \ref{fig:Cl_plot}. The matter power spectrum (left plot) only weakly discriminates between the different cosmologies used here, due to a degeneracy between galaxy bias and the $c_i$ in their effects on the amplitude of the power spectrum. We emphasise that shaded data points are excluded from our analysis, since their correct interpretation requires modelling non-linearities in a way that takes into account non-linear modified gravity effects (which we do not). $f\sigma_8$ constraints from RSDs (right plot), on the other hand, have strong constraining power. In this context we highlight the second and third cosmology, which yield very similar CMB TT power spectra (see figure \ref{fig:Cl_plot}), but very different signatures for $f\sigma_8$. As a result, the third (large $c_M$) cosmology is strongly disfavoured by RSDs.} 
\label{fig:mPk_RSD_plot}
\end{center}
\end{figure}

Constraining power on the $\alpha_i$ primarily comes from three sources: 
\begin{itemize}
\item CMB constraints limit deviations from GR by effectively placing an upper bound on the $\alpha_i$. Large $\alpha_i$ in all parametrisations are generically associated with too much power for the $C_\ell^{TT}$ on large scales (small $\ell$) due to a modified late ISW effect. In the context of the $\alpha_i = c_i \Omega_{\rm DE}$ parametrisation, this was already discussed in \cite{hiclass} and is explicitly shown in figure \ref{fig:Cl_plot} for a number of illustrative choices of the $c_i$ and their corresponding cosmologies. 
\item The onset of gradient instabilities, associated to the stability condition \eqref{gradient_condition_1}, rule out large negative values for both $\alpha_i$. 
\item RSD data further reduce the allowed $\alpha_M$, as can be seen from figure \ref{fig:mPk_RSD_plot}. If $\alpha_M$ is closely correlated/proportional to $\alpha_B$, this results in analogously strong constraints on $\alpha_B$. 
\end{itemize}
For the CMB constraints, note that we used the high-$\ell$, low-$\ell$ {\it and} lensing likelihoods for Planck 2015.\footnote{We note that the addition of the lensing likelihood is crucial to obtain optimal constraints -- we have checked for a fiducial cosmology that pure CMB constraints are significantly weakened without the lensing likelihood.}
Constraints in general are only mildly improved by further adding BAO and mPk data. For BAOs this is due to the fact that we have fixed the background to be $\Lambda{}$CDM. As the angular diameter distance and the Hubble scale (as constrained by the BAO data used here) are background quantities, adding BAO data does not directly add constraining power for the modified gravity $c_i$ parameters. The constraining power of mPk on the $c_i$ is also rather weak. This is because, for the scales considered in our analysis, the $c_{i}$ mainly affect the amplitude of the matter power spectrum, as can be seen from figure \ref{fig:mPk_RSD_plot}, and this effect is degenerate with both galaxy bias and the amplitude of fluctuations. For implementation details see appendix \ref{ap:modeling_details}.

\footnotetext{Specifically, the uncertainties shown correspond to the square root of the diagonal elements of the covariance matrix of the data and thus do not include potential correlations between the data points.}
The addition of RSD constraints primarily affects $c_M$, ruling out large values for this parameter. 
This is because $f\sigma_8$, as constrained by RSD measurements, traces the growth of structure on the associated scales, which strongly depends on the effective strength of gravity. This is predominantly determined by the effective Planck mass, which is increased at late times by larger $c_M$ values. So the growth of structure as measured by the growth function $f$ is significantly more limited when including RSD data than with CMB constraints alone. In addition there is also an additional smaller effect on $c_B$ and effects on $H_0$ and $\sigma_8$, which we will discuss in appendix \ref{ap:extra}. Note that fully propagating modified gravity effects in mapping RSD constraints onto bounds on the $c_i$ (via $f\sigma_8$) is important for extracting the full constraining power of RSDs on the modified gravity parameter space we investigate here (see appendix \ref{ap:modeling_details}). Finally, recall that we use two sets of RSD data, namely samples from the BOSS and 6dF surveys. Interestingly both surveys individually add similar constraining power in terms of the $\alpha_i$ parametrisation parameters, with the BOSS RSD data being marginally more constraining.

\subsection{Robust conclusions}

Having considered the constraints for individual parametrisations above, we would now like to extract those conclusions that are generic and independent of the choice of parametrisation (at least within the representative set of parametrisations we have considered here). 
\\

\nin{\bf Parametrisation-independent conclusions}:  For all parametrisations, $\alpha_M$ can at most be  mildly negative due to gradient instability constraints. This places a tight limit on how much smaller than $M_{\rm Pl}$ the effective Planck mass $M_S$ can be. Depending on the functional form of/parametrisation chosen for $\alpha_M$ this can be strengthened up to ruling out negative $\alpha_M$ altogether (e.g. in parametrisation II). Similarly, while all parametrisations considered above allow mildly negative $\alpha_B$, positive $\alpha_B$ is always preferred at the $~2\sigma$ level (see table \ref{tab:params}).
If $\alpha_M \propto \alpha_B$, then the parametrisations tested suggest that Planck + BAO + mPk constraints generically yield a preferred direction in the associated $c_M, c_B$ plane (see parametrisations I and II in figure \ref{fig:aoParam}), at least for small $\alpha_i/c_i$. This is of interest in the context of models, where this proportionality is a genuine feature of the model and not just an artefact of the parametrisation.
Including RSD data always reduces the allowed parameter range for $\alpha_M$ by ruling out large positive values. This can also further restrict $\alpha_B$, but only if it is sufficiently closely correlated to $\alpha_M$, e.g. if they are proportional to one another (whether this is true is model/parametrisation-dependent).
\\

\nin {\bf Deviations from GR?}: Since negative values of $\alpha_{M,B}$ are strongly constrained by the onset of gradient instabilities, GR occupies a special place in the $\alpha_{M,B}$ plane. In other words, observationally admissible departures from GR are not symmetric in this plane.\footnote{Note that this in fact applies for all theories with $\alpha_M = 0 = \alpha_B$. Also, it is instructive to focus on the $c_i$ in testing `convergence' to GR. In the first two parametrisations considered here this is trivial, but note that for the third parametrisation ($\alpha_i = c_i a^{n_i}$), while $n_i \to \infty$ in a sense recovers GR-like phenomenology for arbitrary $c_i$, it is much cleaner to focus on $c_i \to 0$ as the GR limit of this parametrisation as well.} In all parametrisations $\alpha_M = 0$ provides a good fit to the data. 
When considering only its marginalized constraints, $\alpha_B$ mildly prefers departures from GR at roughly $2\sigma$ confidence level (see table \ref{tab:params}). Note, however, that any statement on model selection needs to take into account both the full parameter space of a given model and its number of degrees of freedom. 
Since a parametrised analysis as performed here is blind to the true number of underlying fundamental parameters and degrees of freedom, we do not perform any model selection analysis in this work.
Finally note that imposing additional theoretical constraints, e.g. as motivated by radiative stability \cite{radstab}, 
has a tendency to eliminate additional non-GR parameter space and therefore tends to drive parameters closer to their GR values. 
So any apparent tension with GR, for a given parametrisation, may at least partially be due to incomplete information about the underlying models and should therefore be interpreted with caution.

\section{Resurrecting $c_{\rm GW} \neq c$ for cosmology}\label{sec:ct}

\begin{figure}[t]
\begin{center}
\includegraphics[width=0.7\linewidth]{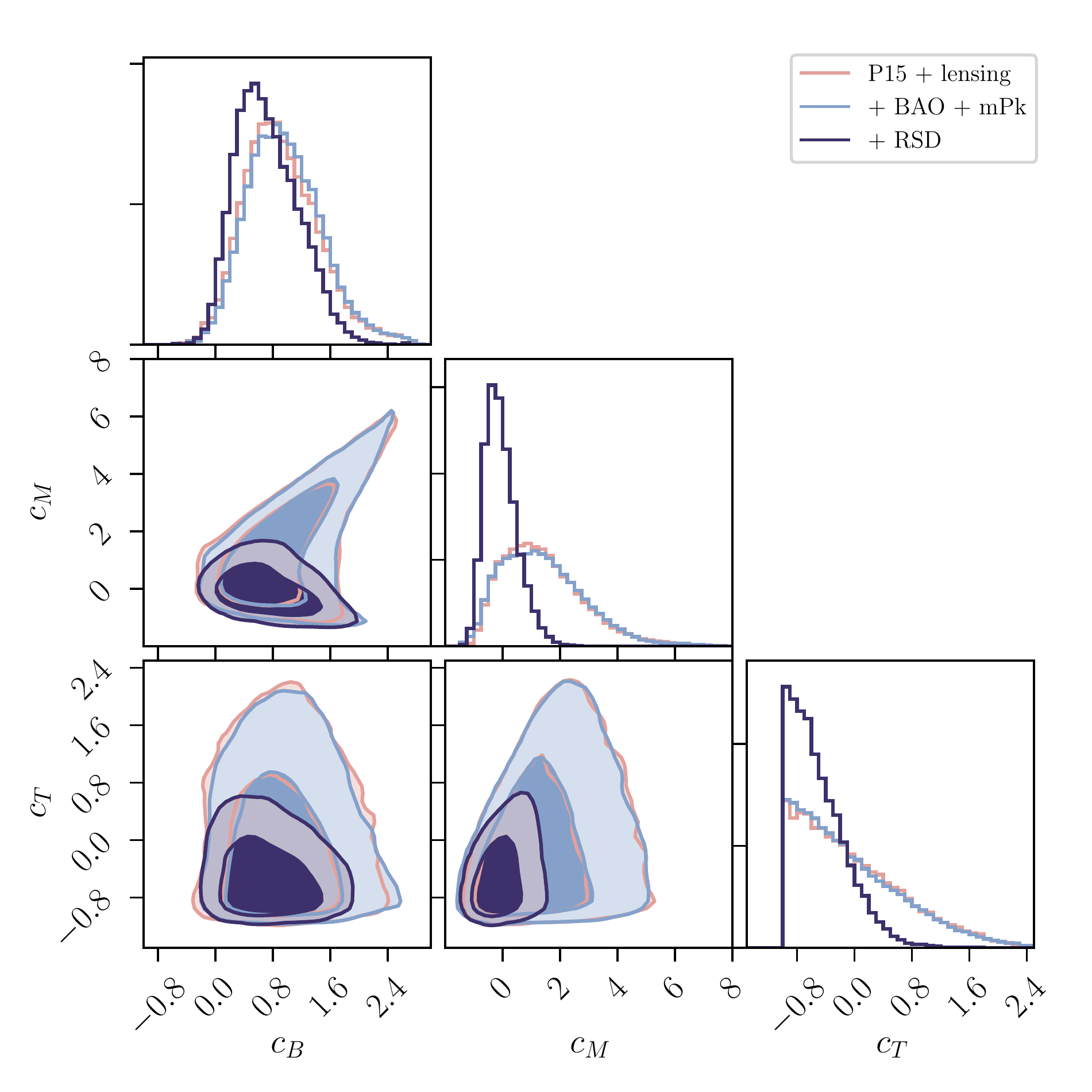}
\end{center}
\caption{Cosmological parameter constraints for the $\alpha_i = c_i \Omega_{DE}$ parametrisation, allowing $\alpha_T$ to vary as well. The inner (outer) contours correspond to 68$\%$ (95$\%$) confidence levels, respectively. Note that constraints in the $c_M, c_B$ plane are very similar to those obtained with a fixed $c_T = 0$ in figure \ref{fig:aoParam} -- see figure \ref{fig:cTComp} for a direct comparison.
Note that, with the addition of RSD measurements, there is a preference for subluminal $c_{\rm GW}$ at the $1.6\sigma$ level -- cf. table \ref{tab:params} and recall that $\alpha_T$ here satisfies $c_{\rm GW}^2 = c^2(1 + \alpha_T) = c^2(1 + c_T \Omega_{DE})$.  
\label{fig:cT}}
\end{figure}

Inferring the equality of the speed of gravitational waves and the speed of light for cosmological energy scales $H_0 \sim 10^{-33} eV$ from the measured equality of those speeds for GW170817 and GRB 170817A (measured at energy scales $\sim 10^{-13} eV$) implicitly assumes a scale/time/energy-independent speed of gravitational waves. As such, this caveat also applies to the derivation of \eqref{Horndeski_simple}. 
Importantly, this means the energy scale probed by GW170817 is significantly larger than that of late-universe cosmology and lies very close to the naive cutoff of theories involving a $G_3$ interaction, usually taken to be $\Lambda_3 = (M_{\rm Pl} H_0^2)^{1/3} \sim 10^{-13} eV$. Making a measurement at those scales therefore in principle tests the (unknown and possibly partial) UV completion of the theory \eqref{Horndeski_simple}, assumed to be governing cosmological dynamics. 
Indeed, as \cite{deRham:2018red} point out, generic Lorentz-invariant UV completions will bring a potentially subluminal cosmological speed of GWs back to luminal for the frequencies observed for GW170817. We refer to \cite{Creminelli:2017sry,deRham:2018red} for a detailed discussion of the naturalness of such a scenario. In any case the large separation in the energy scales of cosmology and those probed by LIGO motivates exploring the cosmology-intrinsic bounds when also varying $c_T$. In this way any conclusion reached does not rely on the properties of a (partial) UV completion of the cosmological theory under consideration. 

\begin{figure}[t]
\begin{center}
\includegraphics[width=0.5\linewidth]{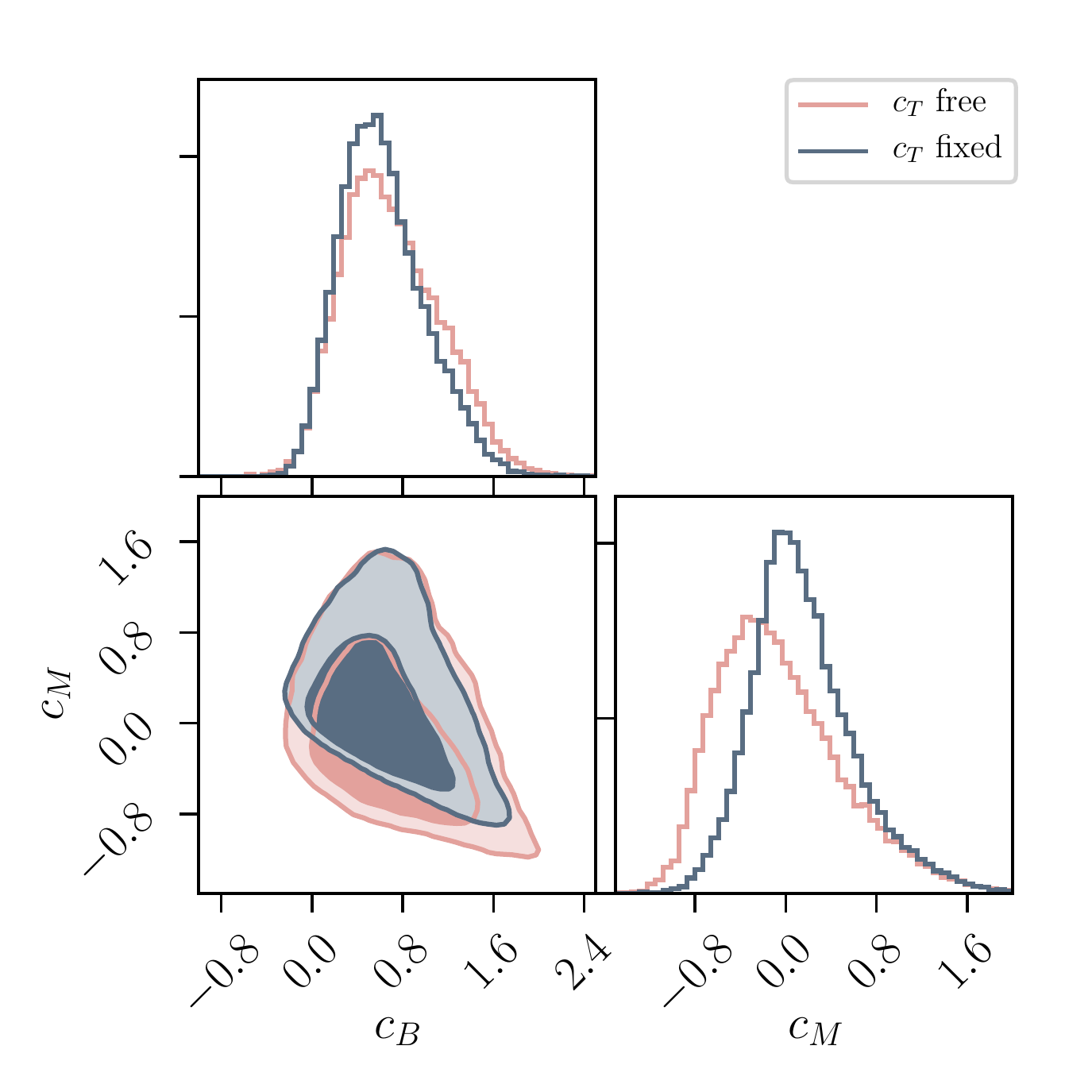}
\end{center}
\caption{Comparison of constraints for $c_M$ and $c_B$ using the $\alpha_i = c_i \Omega_{DE}$ parametrisation, contrasting the case of a fixed luminal speed of gravitational waves ($c_T = 0$) vs. the analogous constraints when that speed is allowed to vary. The inner (outer) contours correspond to 68$\%$ (95$\%$) confidence levels, respectively. Here we use the full P15 + lensing + BAO + mPk + RSD data set. 
Note that, in the $c_M - c_B$ plane, the primary effect of allowing $c_T$ to vary is that the lower boundaries are extended, which is directly related to the fact that a varying $c_T$ affects the onset of gradient instabilities (cf. equation \eqref{gradient_condition_3}) and allows additional cosmologies with mildly negative $c_M$ to avoid developing such instabilities.
\label{fig:cTComp}}
\end{figure}
 
In this section we therefore discuss the constraints from the data sets introduced in section \ref{sec:data} on the full Horndeski theory \eqref{Horndeski_action}, which also allows for a varying $\alpha_T$. We compute constraints for the $\alpha_i = c_i \Omega_{DE}$ parametrisation and impose $c_T \geq -1$ to avoid unphysical, imaginary speeds for gravitational waves (this follows from the gradient (in)stability requirement $\alpha_T \geq -1$). These constraints are shown in Fig.~\ref{fig:cT}. 
First of all notice that contours in the $c_M, c_B$ plane are only mildly changed from the case with fixed $c_T = 0$, presented in figure \ref{fig:aoParam} (a). We explicitly compare these two cases in figure \ref{fig:cTComp}, which shows that the primary difference is due to the fact that a non-zero $c_T$ somewhat shifts the gradient stability condition \eqref{gradient_condition_3}, allowing additional viable cosmologies with mildly negative $c_M$. The overall nature of the constraints on $c_M$ and $c_B$, however, is relatively independent of the (non-)evolution of $c_T$ (at least for this parametrisation). These constraints therefore appear rather robust (under prior changes for $c_T$). 
The addition of RSD data interestingly significantly drives down allowed values of $c_T$, with a preference for negative values and hence sub-luminal propagation of gravitational waves. In this context, note that known bounds on subluminal $c_{\rm GW}$ from the observation of high energy cosmic rays \cite{Moore:2001bv} probe energy scales even larger than LIGO, so should also be ignored in a cosmological context, if the GW170817 constraint is set aside for a cosmological analysis using the above reasoning.\footnote{From this perspective it would also be interesting to re-visit indirect constraints from the energy loss of binary pulsars \cite{Jimenez:2015bwa}.}
In fact, the preference for a subluminal $c_{\rm GW}$ from the data aligns nicely with the theoretical observation, that the existence of a standard UV completion rules out superluminal speeds for the propagation of tensor or scalar perturbations (see e.g. \cite{Adams:2006sv,deRham:2017avq}).

Finally, let us briefly compare the constraints in figure \ref{fig:cT} with those of Ref.~\cite{BelliniParam}, which performed a similar analysis, with two important differences. 
First, \cite{BelliniParam} use a much larger catalogue of RSD measurements than we do here (9 such measurements of $f\sigma_8$ compared to the two used here). We choose this conservative approach in order to safeguard against potential cross-correlations. 
Secondly, \cite{BelliniParam} treat the initial Planck mass as a free parameter, whereas we fix the initial Planck mass to be $M_{\rm Pl}$, as we are interested in late-universe modified gravity/dark energy effects here and do not wish to simultaneously constrain modifications at early times. 
Keeping these two points in mind and comparing the constraints from our figure \ref{fig:cT} with figure 3 of \cite{BelliniParam}, we obtain similar features for $c_B$, but the constraints on $c_M$ differ as \cite{BelliniParam} prefer lower values of $c_M$. However, this is expected for the following two reasons: 
First, as discussed above, RSD data tend to prefer lower values of $c_M$, so it makes sense that adding more RSD measurements and assuming that they are independent strengthens this preference for low $c_M$. 
Note that \cite{BelliniParam} in fact prefer negative values for $c_M$ (and hence a continuously decreasing effective Planck mass in cosmology) at $> 2 \sigma$, whereas there is no such preference for negative $c_M$ for our data sets.
In this context it would be interesting to further investigate potential cross-correlations between the different RSD measurements, as well as possible correlations between RSDs and BAOs.
Secondly, their analysis does not fix the initial value of the effective Planck mass and preferentially samples larger initial values for $M_{\rm Pl}$, which can be partially compensated for by reducing $\alpha_M$. Given the tightly constrained range of allowed initial values for $M_{\rm Pl}$ found by \cite{BelliniParam}, we however expect this second effect to be subdominant. 
Finally, \cite{BelliniParam} also observe a preference of the data for subluminal $c_{\rm GW}$, i.e. negative $c_T$.

\section{Conclusions} \label{sec:conc}

In this paper we have investigated cosmological parameter constraints for general Horndeski scalar-tensor theories, using CMB, redshift space distortion, matter power spectrum and BAO measurements from the Planck, SDSS/BOSS and 6dF surveys. We have focused on computing new constraints for models with luminally propagating gravitational waves (i.e. $c_{\rm GW} = c$ as e.g. motivated by the recent measurements from GW170817 and the assumption of a frequency-independent $c_{\rm GW}$), implementing and discussing several previously unaccounted for aspects in the constraint derivation for such theories. These include a careful handling of stability conditions, restricting the data sets included to safeguard against a potential contamination of results by unaccounted for cross-correlations and  taking into account modified gravity effects on the computation of $f\sigma_8$ (and hence on extracting RSD constraints). 
Together they have strong qualitative effects on the constraints obtained.
Extracting cosmological parameter constraints for any of the above models always requires choosing a parametrisation for the residual functional freedom in such models - at the level of linear cosmology these are the $\alpha_i$ defined in \eqref{alphadef}. To avoid erroneously identifying artefacts of these parametrisations as features of the models to be tested, we compared results for three different parametrisations of the free functions in Horndeski scalar-tensor theories and identified parametrisation-independent features of the constraints. 
The main constraints are shown in figures \ref{fig:aoParam} and \ref{fig:apParam} for three different parametrisations of the $\alpha_i$:  $\alpha_i = c_i \Omega_{DE}$, $\alpha_i = c_i a$ and $\alpha_i = c_i a^{n_i}$, where all $c_i$ and $n_i$ are constants.
Finally, we also investigated models, where $c_{\rm GW}$ is treated as a free function for cosmology (motivated by the fact that the stringent constraints on $c_{\rm GW}$, such as from GW170817, measure this speed at energy scales/frequencies far removed from those relevant for cosmology) and discussed how this affects constraints -- see figures \ref{fig:cT} and \ref{fig:cTComp}.  
Key findings are the following:
\begin{itemize}
\item The running of the Planck mass, $\alpha_M$, is tightly constrained in models where $c_{\rm GW} = c$. Depending on the parametrisation, it can at most be mildly negative, due to strong gradient instabilities that plague models with an effective cosmological Planck mass significantly smaller than $M_{\rm Pl}$. Complementarily, RSD constraints strongly disfavour models with large positive $\alpha_M$. In models where $\alpha_M \propto \alpha_B$, RSD constraints also break degeneracy directions in the associated $c_M - c_B$ plane, exhibited by CMB constraints alone.
$\alpha_B$ is preferentially driven to take positive values in all parametrisations. 
\item CMB constraints are driven by the late ISW effect, with large regions of modified gravity parameter space ruled out by too much power in the TT CMB power spectra on large scales -- see figure \ref{fig:Cl_plot} and notice previous discussions of this effect in \cite{hiclass,Kreisch:2017uet}. RSD measurements are the second main driver of constraints and act via placing tight bounds on $f\sigma_8$, where fully modeling dark energy/modified gravity effects is crucial in order to extract the maximal constraining power. 
\item GR is consistent with the parameter constraints derived here at $\sim 2\sigma$ (see table \ref{tab:params} for parametrisation-specific values). At the level of the `modified gravity functions' $\alpha_i$, any preference for departures from GR is typically driven by the braiding function $\alpha_B$. 
\item For models with $c_{\rm GW} \neq c$ in a cosmological setting (still allowed by GW170817 for a frequency-dependent $c_{\rm GW}$ -- see \cite{deRham:2018red} and the discussion in section \ref{sec:ct}), we show constraints in figure \ref{fig:cT}.
Jointly using CMB and RSD data leads to a $1.6\sigma$ preference for sub-luminally propagating gravitational waves in cosmology  -- cf. related constraints in \cite{BelliniParam}.
\item Constraints on $\alpha_M$ and $\alpha_B$ are mildly affected by freeing up $c_{\rm GW}$ (at least for the parametrisation tested -- see figure \ref{fig:cTComp}) in an interesting way. Due to a modified gradient stability condition, additional viable cosmologies with negative $\alpha_M$ are present in this case.
\end{itemize}

\nin Several future extensions of the work presented here suggest themselves, especially related to the addition of further observational and/or theoretical constraints. 
On the observational front, local constraints e.g. from lunar laser ranging \cite{Williams:2004qba,Babichev:2011iz}, may place additional strong constraints for models that have a sufficiently large cutoff (such that the energy scales tested by such local tests are within the regime of validity of the theory). At larger scales e.g. additional RSD measurements (cf. \cite{BelliniParam}), weak lensing data and galaxy-ISW cross-correlations (cf. \cite{Renk:2016olm,Renk:2017rzu,Lombriser:2016yzn}) promise to add additional constraining power for testing deviations from GR.     
On the theoretical front, e.g. a better understanding of constraints from radiative stability \cite{radstab} and positivity bounds \cite{Adams:2006sv,deRham:2017avq} for gravitational scalar-tensor theories will help in further narrowing down the range of allowed models. By reducing the inherent functional freedom in modified gravity and dark energy theories this will also improve observational bounds on such theories in the process \cite{radstab}.  
The work presented here will hopefully be a useful stepping stone for such future extensions, establishing a number of robust constraints on dark energy and modified gravity models with current data.


\begin{acknowledgments}

We especially thank E. Bellini for numerous discussions and shared insights. We also acknowledge several useful discussions with T. Brinckmann, P. Ferreira, E. Komatsu, C. Kreisch and A. Refregier. 
JN acknowledges support from Dr. Max R\"ossler, the Walter Haefner Foundation and the ETH Zurich Foundation. AN acknowledges support from SNF grant 200021\_169130. In deriving the results of this paper, we have used: CLASS \cite{Blas:2011rf},  corner \cite{corner}, hi\_class \cite{hiclass}, MontePyton \cite{Audren:2012wb,Brinckmann:2018cvx}, xAct \cite{xAct} and xIST \cite{xIST}.

\end{acknowledgments}

\appendix

\section{Theoretical modelling of observations: implementation details}\label{ap:modeling_details}

\nin \textbf{RSD:} Redshift space distortions measure the anisotropic clustering of galaxies in redshift space and are sensitive to the parameter combination $f(z_{\mathrm{eff}})\sigma_{8}(z_{\mathrm{eff}})$, where $f(z_{\mathrm{eff}})$ is the logarithmic linear growth rate, and $\sigma_{8}(z_{\mathrm{eff}})$ is the rms of matter fluctuations in spheres of radius 8 $h^{-1}$ Mpc at the effective redshift of the galaxy sample $z_{\mathrm{eff}}$. In GR, the expression for the logarithmic linear growth rate is given by
\begin{equation}
f(z_{\mathrm{eff}}) = \frac{\mathrm{d}\log{D}}{\mathrm{d}\log{a}},
\end{equation}
where $D$ is the growth factor. 
Within GR, the growth factor $D$ can be estimated using the Heath integral \cite{Heath:1977} for $\Lambda$DCM cosmologies or by solving a GR-specific differential equation for late-time matter perturbations (also valid for a number of minimally coupled dark energy cosmologies -- see e.g. \cite{Dodelson:2003}). These two approaches are not valid for general modified gravity theories and we therefore estimate $f(z_{\mathrm{eff}})$ through
\begin{equation}
\left. f(z_{\mathrm{eff}}) = \frac{\mathrm{d}P^{\sfrac{1}{2}}_{mm, \mathrm{lin}}(k, a)}{\mathrm{d}\log{a}}\right|_{k=k_{\mathrm{fid}}}.
\label{eq:growth_rate}
\end{equation}
We evaluate \ref{eq:growth_rate} at a fiducial wave vector value $k_{\mathrm{fid}}$ using a three-point numerical derivative. As the linear growth rate is defined to be manifestly scale-independent, we choose a fiducial wave vector value such that we can approximate $f(z_{\mathrm{eff}})$ as scale-independent. Specifically, we consider the dependence of $f$ on $k$ at a fixed redshift. As shown in figure \ref{fig:highl_growth_rate} (b), we find the growth rate to be relatively scale-independent for general Horndeski models, except at large scales. We therefore choose a fiducial wave vector well in the scale-independent regime, i.e. $k_{\mathrm{fid}}=0.05$ Mpc$^{-1}$.\footnote{We note that, be default, the growth factor $D$ is computed using the Heath integral within \texttt{hi\_class} and it is thus important to implement the approach outlined above in order to capture all modified gravity effects. We thank Emilio Bellini for pointing this out to us.}
\\

\nin \textbf{mPk:} The matter power spectrum at small scales is affected by nonlinear clustering and potential scale-dependent galaxy bias. Ref.~\cite{Tegmark:2006} model these effects using the fitting formula derived in Ref.~\cite{Cole:2005}. This expression parameterises the relation between linear matter power spectrum and nonlinear galaxy power spectrum, calibrated from N-body simulations. Furthermore, Ref.~\cite{Tegmark:2006} also take into account the BAO smoothing due to nonlinear evolution. In this work, we choose an alternative approach following Ref.~\cite{Dunkley:2009}: as shown in Ref.~\cite{Dunkley:2009}, the smoothing of BAO peaks does not significantly affect the derived constraints on cosmological parameters and we therefore do not include this effect into our analysis. Furthermore, we choose to model the galaxy power spectrum $P_{gg}(k)$ as in Ref.~\cite{Dunkley:2009} i.e.
\begin{equation}
P_{gg}(k) = b^{2}P_{mm}^{\mathrm{nonlin}}(k) + n.
\label{eq:pk_nonlin}
\end{equation} 
The quantity $P_{mm}^{\mathrm{nonlin}}(k)$ is the nonlinear matter power spectrum, $b$ denotes a linear galaxy bias parameter and $n$ parametrises systematic uncertainties due to shot noise and nonlinear evolution. It has been shown in Ref.~\cite{Dunkley:2009} that the model given in Eq.~\ref{eq:pk_nonlin} gives parameter constraints consistent with Ref.~\cite{Tegmark:2006}, while being motivated from perturbation theory. In our work, we model $P_{mm}^{\mathrm{nonlin}}(k)$ using the revised $\textsc{Halofit}$ fitting function \cite{Smith:2003, Takahashi:2012}. We note that $\textsc{Halofit}$ does not include modified gravity effects on the nonlinear matter power spectrum. However, corrections due to nonlinear clustering are small for the wave vector ranges considered in this work and can be captured by the nuisance parameter $n$. As the addition of the matter power spectrum does not significantly modify the derived constraints on modified gravity parameters (they are driven by Planck and RSD data), we thus believe this choice to not affect our conclusions. When estimating constraints on cosmological parameters, we finally marginalise over $b$ and $n$.

\section{Additional constraints and consistency checks}\label{ap:extra}

\begin{figure}[t]
\begin{center}
    \subfloat[Planck\_lite vs Planck\_full]{{\includegraphics[width=0.405\linewidth,valign=b]{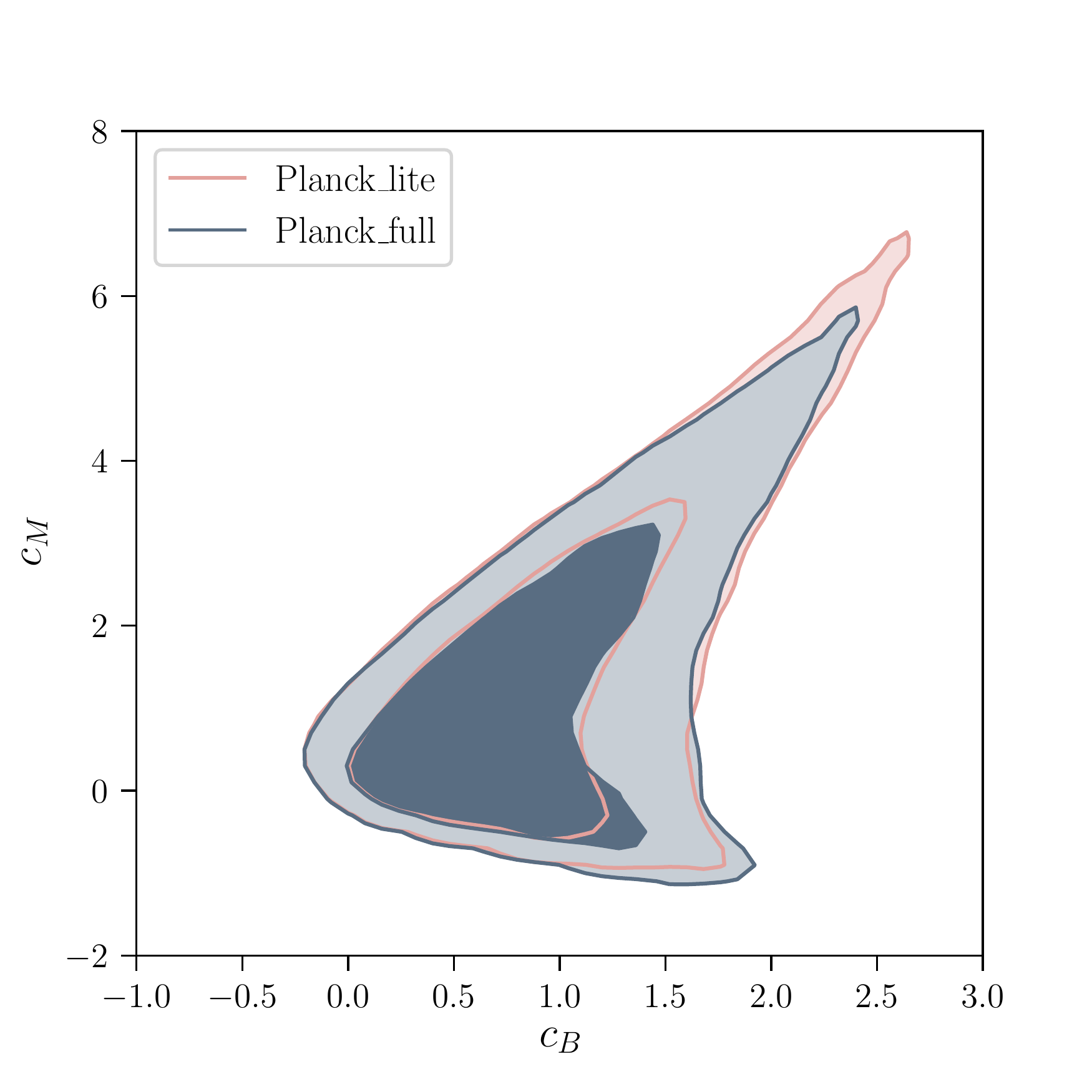} }}%
    \subfloat[The growth rate]{{\includegraphics[width=0.575\linewidth,valign=b]{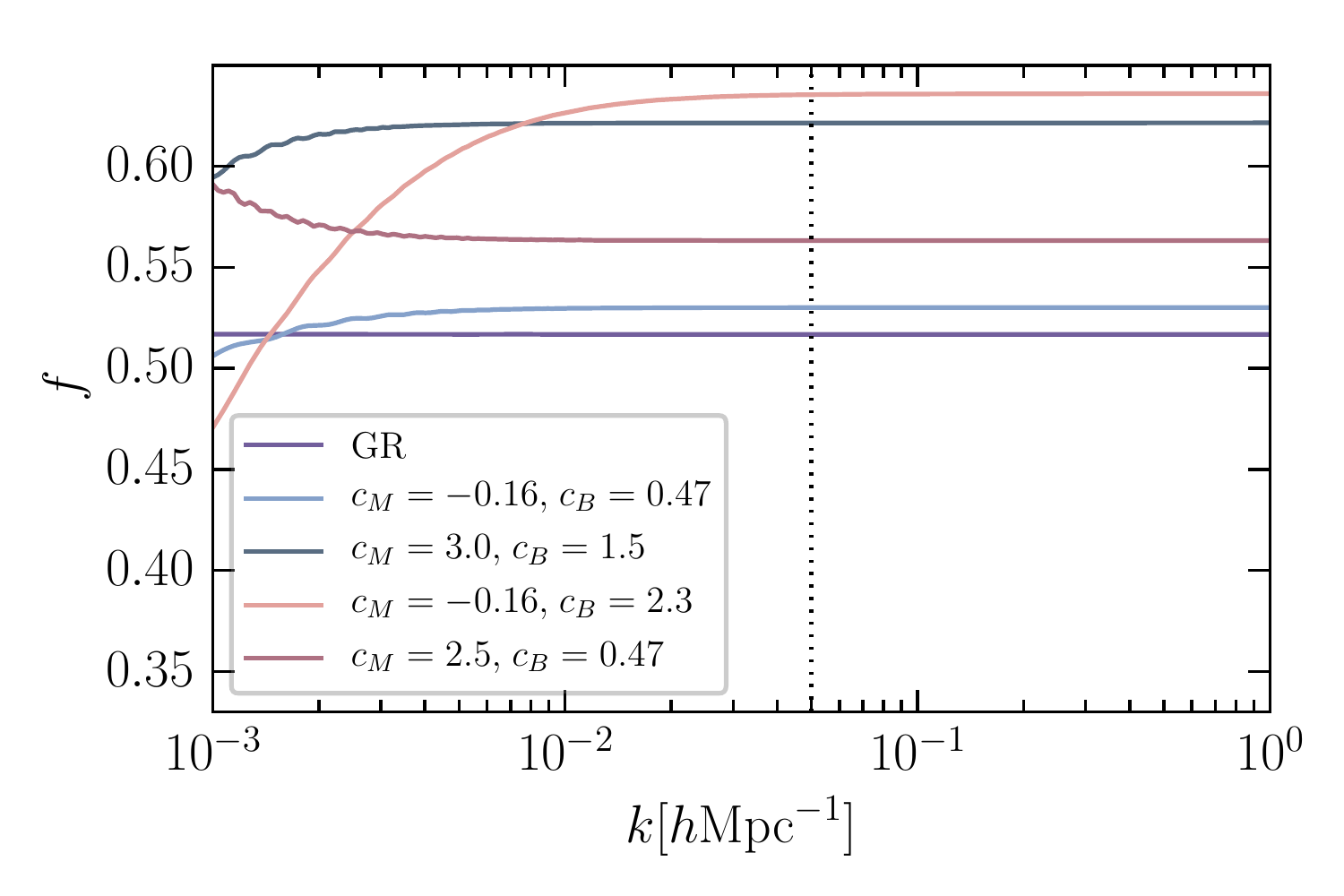}}}%
\caption{(a) Comparison of constraints from P15 + lensing, where the high-$\ell$ likelihood used is either the pre-marginalised high-$\ell$ \texttt{Plik lite} temperature likelihood or the full high-$\ell$ temperature likelihood. We have marginalised over all other cosmological and nuisance parameters here. The two results agree very well. (b) Illustration of the growth rate $f$ as a function of wave vector $k$ for a fiducial cosmological model with various choices of $c_i$ for the parametrisation \eqref{oParam}. The vertical line shows our fiducial wave vector $k_{\mathrm{fid}}=0.05$ Mpc$^{-1}$. We emphasise that the growth rate becomes scale-independent for $k \geq k_{\mathrm{fid}}$ for all cosmologies probed here. }
\label{fig:highl_growth_rate}
\end{center}
\end{figure}

{\bf Planck temperature high-$\ell$ likelihoods}: In figure \ref{fig:highl_growth_rate} (a) we show a comparison of pure Planck 2015 constraints on the $c_i$ parameters of the $\alpha_i = c_i \Omega_{DE}$ parametrisation obtained using the full Planck high-$\ell$ temperature likelihood (with all of its additional nuisance parameters -- 16 in total) vs. the pre-marginalised \texttt{Plik lite} likelihood (which has one nuisance parameter). These two sets of constraints agree very well, justifying the use of the \texttt{Plik lite} likelihood in the derivation of the constraints shown throughout the majority of this paper. Note that the \texttt{Plik lite} likelihood has been pre-marginalised assuming a $\Lambda{}$CDM cosmology, so the fact that we find good agreement is at least partially due to our choice of a $\Lambda{}$CDM background cosmology throughout this paper and, in that sense, unsurprising.  If different background cosmologies are explored, more caution ought to be exercised in using the \texttt{Plik lite} likelihood.
\\

\begin{figure}[t]
\begin{center}
\includegraphics[width=\linewidth]{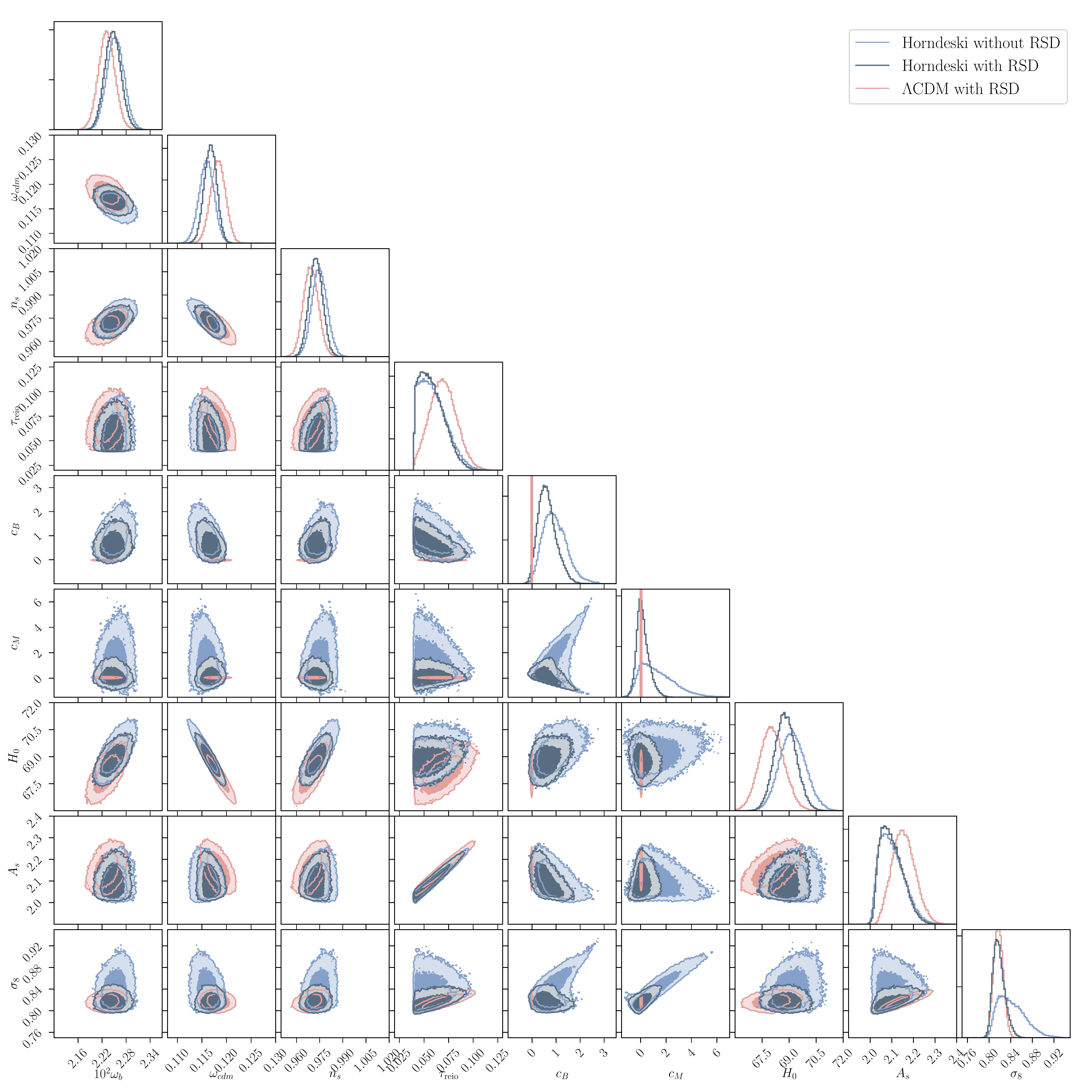}
\end{center}
\caption{Cosmological parameter constraints for all $\Lambda{}$CDM cosmological parameters in addition to the modified gravity/dark energy $c_i$ parameters, using parametrisation \eqref{oParam} for the `Horndeski' contours. For comparison we also show the constraints obtained for a standard $\Lambda{}$CDM model without any additional degrees of freedom. Nuisance parameters are marginalised over and not shown in both cases. As data sets we use P15 + lensing + BAO + mPk data vs. the same data set with additional RSD data, all as described in section \ref{sec:data}. Pure $\Lambda{}$CDM contours are only very mildly affected by the addition of the two RSD measurements we use, so we only plot constraints for the combined data sets in that case. 
\label{fig:oParam_full}}
\end{figure}

\nin{\bf Constraints on other cosmological parameters}: Throughout this paper we have focused on constraints for the modified gravity and dark energy parameters as captured by the $\alpha_i$ and their respective parametrisations. In figure \ref{fig:oParam_full} we now for completeness also show constraints for the standard cosmological $\Lambda{}$CDM parameters for one of the parametrisations used, namely $\alpha_i = c_i \Omega_{DE}$. Here we show constraints obtained using P15 + lensing + BAO + mPk data and constraints obtained once RSD data are added. Figure \ref{fig:aoParam} (a) then is essentially a zoom-in on the $c_M, c_B$ of figure \ref{fig:oParam_full}, so those constraints are of course identical. For comparison we also show constraints on the $\Lambda{}$CDM parameters obtained for a vanilla $\Lambda{}$CDM model without any additional degrees of freedom and for the same data sets. 
In terms of the $\Lambda{}$CDM parameters in modified gravity/dark energy models, the main effect of adding RSD data is on $\sigma_8$ and $H_0$. Both are pushed towards lower values by adding RSD data. In the case of $H_0$ this shift is only mild and still leads to a best-fit $H_0$ larger than in the pure $\Lambda{}$CDM case. For $\sigma_8$ the shift is more notable and one can also notice a `degeneracy direction' in the $c_M - \sigma_8$ plane for CMB + BAO + mPk constraints, that is broken by adding RSD measurements. 
This is because $\sigma_8$ effectively controls the number density of collapsed objects at a given scale. If $c_M$ increases, this means the effective Planck mass at late times increases, so gravity is stronger and objects collapse more efficiently. So it makes sense that these two parameters are correlated. However, while increasing $c_M$ increases $\sigma_8$ at low redshifts ($\sigma_8$ in figure \ref{fig:oParam_full} is measured at redshift zero), at redshifts relevant for CMB constraints different values of $c_M$ have almost no effect on $\sigma_8$. This explains why both $\sigma_8$ measured at redshift zero and $c_M$ are only relatively weakly constrained by CMB (+ BAO + mPk) measurements. Adding RSD data then adds additional and direct sensitivity to the late-universe effects of $c_M$ (and $\sigma_8$), reducing $c_M$ and bringing the posterior for $\sigma_8$ into excellent agreement with the one derived from standard $\Lambda{}$CDM.
Finally, there are also small differences for $\omega_{\rm cdm}$ and $n_s$, which can both be understood in terms of their correlation with $H_0$, i.e. this correlation drives the mild differences in those parameters. 
Note that, motivated by observations of the Gunn-Peterson trough (see e.g. \cite{Becker:2001ee}), we impose a prior $\tau_{\rm reion} \geq 0.04$, corresponding to $z_{\rm reion} \gtrsim 6$. In table \ref{tab:allparam} we furthermore collect parameter constraints for the common parameters varied in all parametrisations for the full P15 + lensing + BAO + mPk + RSD data set.
\\

\begin{figure}[t]
\begin{center}
\includegraphics[width=0.99\textwidth]{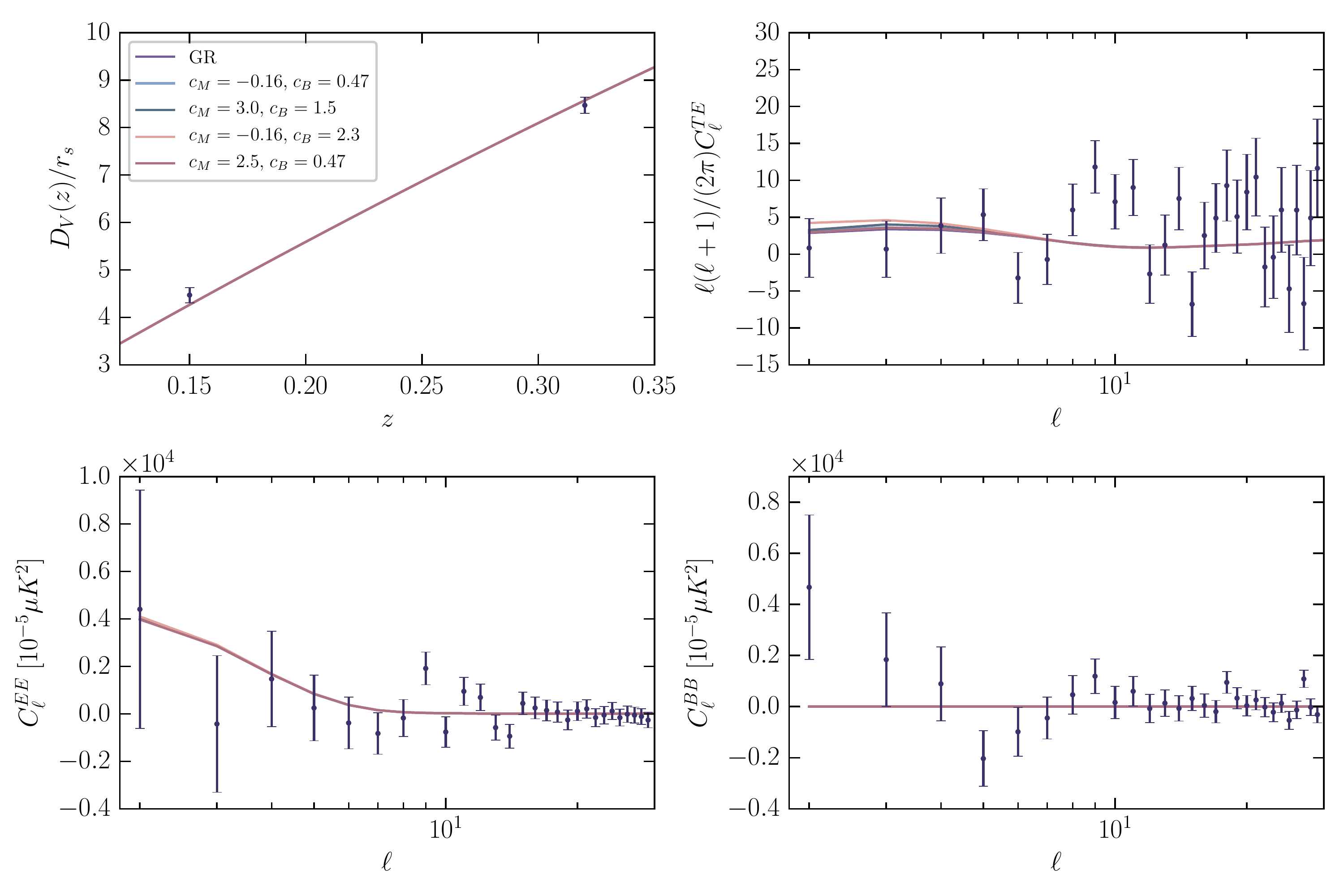}
\caption{Illustration of the BAO (top left) and CMB data used in this work (note that the ${\cal C}_l^{TT}$ are shown in figure \ref{fig:Cl_plot}).  The cosmologies plotted are the same as in figures \ref{fig:Cl_plot} and \ref{fig:mPk_RSD_plot} and the results here should be seen as consistency checks -- as discussed in appendix \ref{ap:extra}, one does not expect there to be any strong signal of modified gravity effects for the observables plotted here.} 
\label{fig:cls-CMB-theor-vs-data}
\end{center}
\end{figure}

\nin{\bf Additional CMB and BAO constraints}: In figure \ref{fig:cls-CMB-theor-vs-data} we show the remaining constraints from BAOs and the other CMB power spectra for the same cosmologies as shown in figures \ref{fig:Cl_plot} and \ref{fig:mPk_RSD_plot} as consistency tests. The BAO measurement of $D_V(z)/r_s$, being a background measurement, unsurprisingly does not discriminate between cosmologies with the same $\Lambda{}$CDM background cosmology (but different perturbations controlled by the $c_i$ parameters). The BB power spectrum is identically zero, since we have set the (primordial) tensor-to-scalar ratio $r = 0$ and the additional scalar modes in Horndeski ST theories do not source tensor/$B$ modes. $E$ modes are also hardly affected at all, so this is a good consistency check that almost all of the CMB constraining power does indeed come from the scalar modes (as thoroughly probed by the $T$ modes), which are modified in the theories we investigate here.

{
\begin{table*}
\begin{center}
\small
\begin{tabular}{cccccc} \hline\hline 
Parameter & Prior & Param. I & Param. II & Param. III & Param. I*\\ \hline \Tstrut                             
$100\theta_{\mathrm{s}}$ & flat unbound & $2.25 \pm 0.04$ & $2.24 \pm 0.04$ & $2.24 \pm 0.04$ & $2.25 \pm 0.04$\\ 
$w_{\mathrm{cdm}}$ & flat unbound & $0.117\substack{+0.002 \\ -0.003}$ & $0.117 \pm 0.003$ & $0.117 \pm 0.003$ & $0.117 \pm 0.003$\\
$w_{\mathrm{b}}$ & flat unbound & $0.0104 \pm 0.0008$ & $0.0104 \pm 0.0008$ & $0.0104 \pm 0.0008$ & $0.0104 \pm 0.0008$\\
$n_{\mathrm{s}}$ & flat unbound & $0.973 \pm 0.009$ & $0.972 \pm 0.009$ & $0.972 \pm 0.009$ & $0.973 \pm 0.009$\\
$\log{10^{10}A_{\mathrm{s}}}$ & flat unbound & $3.05\substack{+0.053 \\ -0.037}$ & $3.05\substack{+0.06 \\ -0.05}$ & $3.06 \pm 0.05$ & $3.04\substack{+0.05 \\ -0.04}$\\
$\tau_{\mathrm{reion}}$ & flat $\in [0.04, \,-]$ & $0.0588\substack{+0.0276 \\ -0.0177}$ & $0.0662\substack{+0.0302 \\ -0.0239}$ & $0.0677\substack{+0.0267 \\ -0.0237}$ & $0.0588\substack{+0.0280 \\ -0.0177}$\\ \\ \hline \\
$c_{M}/c_{\delta M}$ & flat unbound & $0.20\substack{+1.15 \\ -0.82}$ & $0.27\substack{+0.54 \\ -0.26}$ & $0.30\substack{+0.77 \\ -0.45}$ & $-0.01\substack{+1.30 \\ -0.87}$\\
$c_{B}$ & flat unbound & $0.63\substack{+0.83 \\ -0.62}$ & $0.48\substack{+0.83 \\ -0.46}$ & $1.1\substack{+0.89 \\ -1.10}$ & $0.71\substack{+0.88 \\ -0.71}$\\ \\ \hline \\
$A_{\mathrm{Planck}}$ & flat $\in [0.9, \,1.1]$ & $1.00 \pm 0.5$ & $1.00 \pm 0.5$ & $1.00 \pm 0.5$ & $1.00 \pm 0.5$\\
$b$ & flat $\in [1., \,3.]$ & $2.19 \pm 0.06$ & $2.16 \substack{+0.10 \\ -0.12}$ & $2.19 \substack{+0.09 \\ -0.11}$ & $2.22 \substack{+0.11 \\ -0.13}$\\
$n$ & flat $\in [0., \,21390.]$ & $1494.\substack{+3000. \\ -1430.}$ & $1427.\substack{+2764. \\ -1360.}$ & $1434.\substack{+2862. \\ -1374.}$ & $1477.\substack{+2913. \\ -1415.}$\\
\hline \hline
\end{tabular}
\end{center}
\caption{Common parameters varied in the MCMC analysis for the different parametrisations used in this paper with their respective priors and posteriors.  The uncertainties shown denote the $95 \%$ confidence level. $A_{\mathrm{Planck}}, b$ and $n$ are nuisance parameters. Param I* refers to parametrisation I with a varying $c_T$. The $c_M/c_{\delta M}$ row shows posteriors  for $c_{\delta M}$ for parametrisation III and for $c_M$ for all other parametrisations.}
\label{tab:allparam}
\end{table*}
}

\bibliographystyle{JHEP}
\bibliography{HorndeskiMCMC}

\end{document}